# Investigation of Systematic Bias in Radiometric Diameter Determination of Near-Earth Asteroids: the Night Emission Simulated Thermal Model (NESTM)

Stephen D. Wolters[a] and Simon F. Green[a]

[a] *Planetary and Space Sciences Research Institute, The Open University, Walton Hall, Milton Keynes, MK7 6AA, UK*

Please direct editorial correspondence and proofs to:

Stephen Wolters
Planetary and Space Sciences Research Institute
The Open University
Walton Hall
Milton Keynes
Buckinghamshire
MK7 6AA
UK

Phone: +44 1908 659465

e-mail: s.d.wolters@open.ac.uk

e-mail addresses of co-author: s.f.green@open.ac.uk

## Abstract

The Near-Earth Asteroid Thermal Model (NEATM, Harris, 1998) has proven to be a reliable simple thermal model for radiometric diameter determination. However NEATM assumes zero thermal emission on the night side of an asteroid. We investigate how this assumption affects the best-fit beaming parameter $\eta$, overestimates the effective diameter $D_{eff}$ and underestimates the albedo $p_v$ at large phase angles, by testing NEATM on thermal IR fluxes generated from simulated asteroid surfaces with different thermal inertia $\Gamma$. We compare NEATM to radar diameters and find that NEATM overestimates the diameter when $\eta$ is fitted to multi-wavelength observations and underestimates the diameter when default $\eta$ is used. The Night Emission Simulated Thermal Model (NESTM) is introduced. NESTM models the night side temperature ($T_{night}$) as an iso-latitudinal fraction ($f$) of the maximum day side temperature ($T_{max}$ calculated for NEATM with $\eta = 1$): $T_{night} = fT_{max} \cos^{1/4} \phi$, where $\phi$ is the latitude. A range of $f$ is found for different thermal parameters, which depend on $\Gamma$. NESTM diameters are compared with NEATM and radar diameters, and it is shown that NESTM may reduce the systematic bias in overestimating diameters. It is suggested that a version of the NESTM which assumes $\Gamma = 200$ J m$^{-2}$ s$^{-1/2}$ K$^{-1}$ is adopted as a default model when the solar phase angle is greater than 45°.

# 1   Introduction

## 1.1   Using Simple Thermal Models for Asteroid Diameter Determination

Simple thermal models are a useful way of determining the diameter and albedo of asteroids in the case where there is limited knowledge of their physical properties, particularly shape and rotation axis, and where mid infrared spectra or broadband fluxes have only been obtained from very few aspects. For most asteroids where thermal IR spectra have been obtained this case holds, and hence simple thermal models remain important in using this information to derive asteroid size distributions.

A thermophysical model [e.g. Harris et al. (2005), Müller et al. (2005)] can provide more accurate results than simpler thermal models if accurate measurements of the thermal continuum emission over a range of aspect angles are available, together with knowledge of parameters such as shape and pole orientation. Unfortunately these conditions hold for only a few NEAs observed to date, and the situation is unlikely to change in the near future. Also, it might not be practicable to obtain them for many objects in any future thermal IR survey. Therefore we need to rely on simple models, just as IRAS (Tedesco 1992) used the Standard Thermal Model in the asteroid main belt asteroids (Lebofsky & Spencer 1989, and references therein). For NEAs, the Near-Earth Asteroid Thermal Model (NEATM, Harris 1998) has proven to be the most reliable. However NEATM assumes zero thermal emission on the night side of an asteroid, which may lead to significant overestimation of diameter and underestimation of albedo. This affects the beaming parameter $\eta$ and contributes to a trend of increasing $\eta$ with phase angle $\alpha$. In this paper we investigate whether a simple thermal model that assumes non-zero night side emission can improve on NEATM.

## 1.2   The Need for Improved Physical Characterisation of NEAs

Improved physical characterisation of Near-Earth Asteroids (NEAs) is important for understanding their origin and evolution, the links between meteorites and their parent bodies, and for assessing the impact hazard. NEAs are also representative of small main belt asteroids (Binzel *et al.* 2002). Cellino *et al.* (2002) describe how the discovery rate of NEAs is vastly outstripping their investigation. As of November 2008, the number of NEAs with measured diameters and albedos is about 87 (http://earn.dlr.de/nea/table1_new.html) while the total number of NEAs discovered is over 5800 (http://neo.jpl.nasa.gov/).

Improved statistics of the diameters and albedos of NEAs are needed for a more accurate derivation of their size distribution, which is crucial for assessment of the impact hazard and for optimising survey strategies. "All estimates of NEA populations and impact rates are plagued by uncertainties concerning the albedos of NEAs and hence by the conversions from the observed quantities (magnitudes) to size and energy" (Morrison 2008). Smaller NEAs below 1 km particularly need to be characterized; but unfortunately there is a bias against selecting small, low albedo objects, and succeeding in observing small, high albedo objects at mid-IR wavelengths. As the number of NEAs with known taxonomic type increases, so does the requirement for an increase in measurements of their albedos. If an albedo distribution is derived for each taxonomic type it can be used to derive a de-biased size distribution. Stuart and Binzel (2004) have done the first study using albedo statistics from NEAs, obtained from Delbó *et al.* (2003), which relies on simple thermal models. However, A, R and U-types are still obtained from main-belt statistics and several values are based on very few classified objects (for example the D-type complex has one member with a measured albedo).

Also, trends within taxonomic types may reveal surface processes. The majority of NEAs with measured albedos are S-types. Delbó (2004) has found a trend of increasing albedo with decreasing size among S-types and interprets it as evidence for space weathering. Wolters *et al.* (2008) found a possible trend of increasing beaming parameter with diameter for small S- and Q-type asteroids. As the available data for other taxonomic types grow, there may be other similar trends discovered.

## 1.3    The Radiometric Method of Diameter Determination

Radiometric diameter determination is a powerful method for acquiring the diameters of a large number of objects, and therefore would be a suitable method for a survey to improve the statistics of physically characterised NEAs. Thermal models can be fitted to thermal IR fluxes to derive the size and albedo of an asteroid. The size is ultimately presented as the effective diameter $D_{eff}$, the equivalent diameter of a perfect sphere with the same projected area as the (generally) irregularly shaped asteroid. The albedo is presented as the geometric albedo $p_v$, the ratio of the visual brightness to that of a perfectly diffusing 'Lambertian' disk of the same diameter. The bolometric Bond albedo $A$ can be related to $p_v$ through:

$$p_v = \frac{A}{q} \qquad (1)$$

where $q$ is the phase integral (Bowell $et\ al.$ 1989). For a given absolute visual magnitude $H_V$, there is a range of possible $p_v$ and hence $D_{eff}$ described by (e.g. Fowler & Chillemi 1992):

$$D_{eff}\,(\mathrm{km}) = \frac{10^{-H_v/5}\,1329}{\sqrt{p_v}} \qquad (2)$$

The principle of the radiometric method is described in Morrison (1973) and Lebofsky and Spencer (1989), with more recent reviews by Delbó and Harris (2002) and Harris and Lagerros (2002). The energy balance depends on the projected area and the albedo. Since the reflected solar component is proportional to $A$ and the thermal component is proportional to (1-$A$), simultaneous measurements of both can provide a unique $D_{eff}$ and $p_v$ via the radiometric method of diameter determination.

## 1.4    The Near-Earth Asteroid Thermal Model (NEATM)

The NEATM is a modification of the so-called Standard Thermal Model (STM) which, as outlined in Lebofsky $et\ al.$ (1986), considers the asteroid as a spherical non-rotating object, with a surface temperature in instantaneous equilibrium with incoming solar radiation, with a temperature distribution decreasing from a maximum at the subsolar point ($T_{max}$) to zero at the terminator, and no thermal emission on the night side. The beaming parameter $\eta$ was introduced to take account of enhanced sunward thermal emission due to the surface roughness. In the STM $\eta$ = 0.756, calibrated from the occultation diameters of (1) Ceres and (2) Pallas, and $T_{max}$ becomes:

$$T_{\text{fit}} = \left[ \frac{(1-A)S_0}{r^2 \varepsilon \sigma \eta} \right]^{\frac{1}{4}}$$

(3)

where $S_0$ = 1374 W m$^{-2}$ is the solar flux at 1 AU, $r$ =distance from the Sun in AU, $\varepsilon$ = emissivity (assumed $\varepsilon$ = 0.9), and $\sigma$ = Stefan-Boltzmann constant.

Harris (1998) introduced the NEATM as an appropriate model for NEAs, which are often observed at high phase angle and are thought to have higher surface thermal inertia, as a consequence of their smaller size in comparison to observable main-belt asteroids (smaller asteroids may have higher rotation rates and less regolith). NEATM modifies the STM in two ways. First, it allows $\eta$ in Eq. 3 to be varied until the model fluxes $F_{mod}(n)$ give a best fit to the observed thermal IR spectrum $F_{obs}(n)$, effectively forcing the model temperature distribution to show a colour temperature consistent with the apparent colour temperature implied by the data. Second, it replaces the STM phase angle correction in the same way as the projected model (e.g. Cruikshank & Jones 1977), which models the asteroid as a sphere and calculates the temperature on the surface assuming Lambertian emission and zero emission on the night side. The projected model is the equivalent of the NEATM with $\eta$ = 1 (i.e. with no beaming).

The NEATM temperature distribution is defined by the longitude $\theta$ and latitude $\phi$ on the asteroid surface, where $\theta = 0°$ and $\phi = 0°$ are at the subsolar point.

$$T(\theta,\phi) = 0 \text{ for } \frac{\pi}{2} \leq \theta \leq \frac{3\pi}{2}$$

$$T(\theta,\phi) = T_{fit} \cos^{\frac{1}{4}}\theta \cos^{\frac{1}{4}}\phi \text{ for } -\frac{\pi}{2} < \theta < +\frac{\pi}{2} \text{ and for } -\frac{\pi}{2} \leq \phi \leq +\frac{\pi}{2} \quad (4)$$

The model fluxes $F_{mod}(n)$ are calculated by integrating $B(\lambda_n, T(\theta,\phi))$ over the portion of the asteroid surface visible to the observer:

$$F_{mod}(n) = \frac{\varepsilon D_{eff}^2}{4\Delta^2} \int_{-\frac{\pi}{2}}^{+\frac{\pi}{2}} \int_{\alpha-\frac{\pi}{2}}^{+\frac{\pi}{2}} B(\lambda_n, T(\theta,\phi)) \cos^2\phi \cos(\alpha-\theta) \mathrm{d}\theta \, \mathrm{d}\phi \quad (5)$$

where $\Delta$ = distance to the Earth in AU.

Finding an accurate $\eta$ requires good wavelength sampling of the thermal continuum, ideally at least four filter measurements over the range 5-20 µm (e.g. Delbó et al. 2003), although $\eta$-fitting for observations over a narrower range (8-12.5 µm) but with higher spectral resolution has proved successful (e.g. Wolters *et al.* 2005, 2008). Delbó *et al.* (2003) found a trend of increasing beaming parameter $\eta$ with phase angle $\alpha$. From this trend, they proposed a default $\eta = 1.0$ for observations α < 45° (equivalent to projected model) and $\eta = 1.5$ for α ≥ 45°, for the case where only one or two N- and/or Q-band observations are available, or the spectral resolution is not high enough to make $\eta$-fitting sensible.

Fitting $\eta$ compensates somewhat for effects from surface thermal inertia, surface roughness and other deviations from the STM. However, NEATM ignores thermal emission on the night side and, even with $\eta$-fitting, the resulting diameter is overestimated and consequently the albedo is underestimated, increasingly at higher phase angles. This also contributes to the trend of increasing $\eta$ with $\alpha$. If this systematic error was large, one might expect a trend of decreasing albedo with phase angle. Delbó (2004) did not find a trend and suggested that this indicated that such an error did not play a major role up to α ~ 60°. However, we note that the default model uncertainty in the measurement of $p_v$ is 30% and the sample size is still small. Delbó (2004) also compared NEATM diameters to radar diameters in a similar analysis to our own in Section 3.2, and found that there was a systematic bias in overestimating diameters of 8% ± 4%, which could be explained as a consequence of ignoring thermal emission on the night side.

## 1.5    The Fast Rotating Model (FRM)

Previously in situations where asteroids were observed at high phase angles and the NEATM often gave poor fits, it has been necessary to resort to the Fast Rotating Model (FRM, Lebofsky & Spencer 1989) to derive an asteroid's diameter and albedo, for example: 1999 NC$_{43}$ and 2002 BM$_{26}$ in Delbó et al. (2003), 1999 HF$_1$ in Wolters et al. (2005). In the FRM, the temperature contours of an assumed spherical asteroid, with a rotation axis at 90° to the solar direction, are smoothed out due to a combination of thermal lag and rotation which causes received solar flux at a given latitude $\phi$ to be re-emitted at a constant rate, without cooling as it rotates. Consequently the temperature distribution depends only on latitude, and the day and night side are at an equal temperature. The FRM can be regarded as the opposite extreme to the STM.

## 1.6    The Modified Projected Model

The modified projected model was introduced by Green *et al.* (1985) as an appropriate model to fit to thermal IR fluxes of NEA (3200) Phaethon, which was observed at a reasonably high phase angle (48°), and for which the STM fit badly. The model is considered in more detail in Green (1985).

Whereas the NEATM assumes that there is no night side emission, the modified projected model uses a parameter $f$ to define the night side temperature, so that for a latitude $\phi$ the night side temperature is:

$$T_{\mathrm{night}} = fT_{\mathrm{max}} \cos^{\frac{1}{4}} \phi \qquad (6)$$

where $T_{max}$ is defined as in Eq. 3 without $\eta$, i.e. beaming is not considered in this model. Setting $f = 0$ would be the equivalent of the projected model, which is itself the equivalent of the NEATM with the beaming parameter $\eta = 1$. In order to conserve energy, $T_{max}$ is replaced by a reduced maximum day side temperature $T_{mod}$. The day side temperature for a given latitude $\phi$ and longitude $\theta$ is given by:

$$T_{day} = T_{mod} \cos^{\frac{1}{4}} \theta \cos^{\frac{1}{4}} \phi \tag{7}$$

For a particular latitude and longitude on the day side, if $T_{night}$ is greater than $T_{day}$ then the night side temperature takes precedence. $T_{mod}$ is calculated by balancing the total emitted flux to that absorbed in the energy balance equation:

$$\frac{\pi D_{eff}^2 (1 - A)}{4r^2} = \frac{D_{eff}^2}{2} \varepsilon \sigma \int_{\frac{-\pi}{2}}^{\frac{\pi}{2}} \int_0^\pi G\left( T_{mod}^4 \cos\theta, f^4 T_{max}^4 \right) \cos^2\phi \, d\theta \, d\phi \tag{8}$$

where $T_{mod} < T_{max}$, and $G(x, y) = x$ if $x > y$ and $G(x, y) = y$ if $x < y$, which can then be solved iteratively to give $T_{mod}$.

The emitted flux measured from Earth (outside the atmosphere) is calculated by integrating over the visible hemisphere longitudes and latitudes using the black-body function for each surface element:

$$
\begin{aligned}
F_{mod}(n) = \frac{D_{eff}^2}{2\Delta^2} \varepsilon \int_{\frac{-\pi}{2}}^{\frac{\pi}{2}} &\left[ \int_{\alpha - \frac{\pi}{2}}^{\frac{\pi}{2}} B\left( \lambda(n), G\left( T_{mod} \cos^{\frac{1}{4}}\theta, f T_{max} \right) \cos^{\frac{1}{4}}\phi \right) \cos(\alpha - \theta) d\theta \right. \\
&\left. + \int_{\frac{\pi}{2}}^{\alpha + \frac{\pi}{2}} B\left( \lambda(n), f T_{max} \cos^{\frac{1}{4}}\phi \right) \cos(\alpha - \theta) d\theta \right] \cos^2\phi \, d\phi
\end{aligned}
\tag{9}
$$

$f$ depends on the asteroid's spin axis, rotation period, thermal inertia and shape. Green (1985) was able to vary $f$ to provide a best-fit to Phaethon ($f = 0.65 \pm 0.02$).

Hansen (1977) has also discussed using a non-zero night side temperature distribution. Hansen ensured a smooth transition from the day side to the night side temperature by introducing a monotonically increasing function $f(\theta)$ which is 0 for $\theta = 0°$ and 0.60 for $\theta \geq 90°$. In Hansen's model the maximum day side temperature is not recalculated to conserve energy.

## 2  The Model

### 2.1  Combining NEATM with the Modified Projected Model

The Night Emission Simulated Thermal Model (NESTM) combines the NEATM with features of the modified projected model. Additionally to the parameters required to run NEATM, an assumed surface thermal inertia $\Gamma$ and rotation period $P$ are also needed. If $P$ is not known, we assume $P = 5$h, which is the average for NEAs (Binzel et al. 2002). Hence we can define different versions of NESTM, depending on the assumed $\Gamma$. In this paper we consider NESTM with $\Gamma = 40, 120, 200, 550,$ and 2200 J m$^{-2}$ s$^{-1/2}$ K$^{-1}$ which we refer to as NESTM40, NESTM120, NESTM200, NESTM550 and NESTM2200 respectively. $\Gamma = 40$ represents a "dusty" $\Gamma$ approximately equivalent to that of the lunar surface. $\Gamma = 200 \pm 40$ J m$^{-2}$ s$^{-1/2}$ K$^{-1}$ has been found to be the average for NEAs (Delbo *et al.* 2007) based on a statistical inversion study of asteroids between 0.8 and 3.4 km diameter. Delbó (2004) found that the average NEA $\Gamma$ from a study of evening/morning effect limiting curves on $\alpha$-$\eta$ plots was $\Gamma = 550 \pm 100$ J m$^{-2}$ s$^{-1/2}$ K$^{-1}$. $\Gamma = 2200$ J m$^{-2}$ s$^{-1/2}$ K$^{-1}$ represents a "bare rock" surface equivalent to that of granite (NB thermal inertia is a function of temperature $\Gamma \propto T^{3/2}$ so a planetary body with the same surface will have a lower surface thermal inertia if it orbits further from the Sun).

The degree to which the surface of an asteroid responds to changes in insolation can be characterised by the thermal parameter $\Theta$, which combines the rotation rate $\omega = 2\pi/P$, the surface thermal inertia $\Gamma$ and the STM maximum temperature $T_{max}$ (Spencer *et al.* 1989):

$$\Theta = \frac{\Gamma\sqrt{\omega}}{\varepsilon\sigma T_{\max}^{3}} \qquad (10)$$

NESTM applies an iso-latitudinal night side temperature (Eq. 6) that is a fraction $f$ of the maximum day side temperature when $\eta = 1$ ($T_{max}$). We have to use an average night side temperature for each latitude, because for most objects we do not know the pole orientation and hence the rotation direction. Hence, we do not know if we are observing the warmer "afternoon" side, which we refer to as positive phase angle (+$\alpha$), or the cooler "morning" side (-$\alpha$). So whereas NEATM always has too low thermal emission on the night side, with NESTM it can be higher or lower, albeit on average closer. It can be considered that NESTM applies a "damped down" version of the FRM to the night side of the asteroid.

The $f$ parameter is a function of $\Theta$. Clearly, since $\Theta$ is dependent on $T_{max}$, which is in turn dependent on the albedo, we must recalculate the asteroid's thermal parameter for every $p_v$, so the model is run with a look-up table with an appropriate $f$ for any given small range of thermal parameter. We describe how it was generated in Section 2.2. $\Theta$ is much more strongly dependent on $\Gamma$ and $P$ than $p_v$, such that the appropriate $f$ does not typically change by more than 0.02 as a range of $p_v$ is run through, and so it would be an acceptable simplification to run the model with a fixed $f$ parameter for the whole range of $p_v$ if required.

The beaming parameter $\eta$ is applied to the day side, so that day side temperatures $T_{day}$ are calculated as described by Eqs. 3 and 4 for the NEATM. The major departure from the modified projected model here is that a modified maximum day side temperature $T_{mod}$ is no longer iteratively calculated using the energy balance Eq. 8. Instead the beaming parameter $\eta$ is fitted in the same manner as for NEATM, effectively measuring the real day side temperature from the observed thermal IR fluxes. This makes the model considerably simpler than the modified projected model. Like the modified projected model, if $T_{night} > T_{day}$ at any point on the day side then $T_{night}$ is used. The modelled flux $F_{mod}(n)$ is determined from the temperature array by integrating the black body function over the visible surface (c.f. Eq. 9 where not all visible longitudes were integrated, since the night side emission was assumed to be zero):

$$F_{mod}(n) = \frac{D_{eff}^2}{2\Delta^2}\varepsilon\int_{-\frac{\pi}{2}}^{\frac{\pi}{2}}\left[\int_{\alpha-\frac{\pi}{2}}^{\frac{\pi}{2}}B\left(\lambda(n), G\left(T_{fit}\cos^{\frac{1}{4}}\theta, fT_{max}\right)\cos^{\frac{1}{4}}\phi\right)\cos(\alpha-\theta)\mathrm{d}\theta\right.$$
$$\left.+\int_{\frac{\pi}{2}}^{\alpha+\frac{\pi}{2}}B\left(\lambda(n), fT_{max}\cos^{\frac{1}{4}}\phi\right)\cos(\alpha-\theta)d\theta\right]\cos^2\phi\,\mathrm{d}\varphi$$

$(11)$

where $G(x, y) = x$ if $x>y$ and $G(x, y) = y$ if $x<y$.

## 2.2    Defining an Appropriate $f$ Parameter

Applying a night side constant temperature profile as a latitude-dependent fraction $f$ of $T_{max}$ is just an approximation of the effect that a body with significant thermal inertia would have on the temperature profile. In reality the temperature on the night side would slowly cool from the day side temperature. We can model the temperature for the night side for an asteroid with a given thermal inertia $\Gamma$, rotation period $P$, albedo $A$, and heliocentric distance $r$ using a simple one dimensional thermophysical model, applying essentially the same method as Wesselink (1948) (see also Wolters, 2005). The thermophysical model was run for an asteroid with bolometric Bond albedo $A = 0.2$ at a distance from the Sun of $r = 1$ AU and at thermal inertias $\Gamma = 40, 100, 550$ and $2200$ J m$^{-2}$ s$^{-1/2}$ K$^{-1}$. It was run for rotation periods 1-100 h. The model assumes that the pole orientation is 90° and that the asteroid is spherical.

The average equatorial night side temperature $\overline{T}$ (90° $\geq \theta \geq$ 270°) is found for each model run, from which we can derive $f$:

$$f = \frac{\overline{T}\left(90° \geq \theta \geq 270°, \phi = 0\right)}{T_{max}} \qquad (12)$$

We can relate these $f$ parameters to the thermal parameter $\Theta$ through Eq. 10, and hence have an appropriate value to use for any heliocentric distance, rotation period and albedo in the NESTM. The variation of $f$ with $\Theta$ is given in Fig. 1, and the resulting look-up table is given in Table 1.

## 3    Assessing the NESTM

### 3.1    Testing on Simulated Surfaces

In order to assess whether the NESTM measures diameters more accurately than the NEATM, a set of temperature distributions of a test asteroid were created using the thermophysical model. We used an asteroid with parameters $p_v$ = 0.25, $r$ = 1 AU, $P$ = 5 h, and $\Gamma$ = 40, 120, 200, 550 and 2200 J m$^{-2}$ s$^{-1/2}$ K$^{-1}$. Assigning a slope parameter $G$ = 0.15, the asteroid's albedo is equivalent to $A$ = 0.09815. Following Eq. 10, the asteroid's surface has thermal parameter $\Theta$ = 0.238, 0.714, 1.190, 3.273 and 13.094 respectively. The resulting temperature distribution generated with the thermophysical model for $\Gamma$ = 200 J m$^{-2}$ s$^{-1/2}$ K$^{-1}$ and the equatorial temperatures for all five cases are shown in Fig. 2. To contrast the different models, the NESTM200 and NEATM temperature distributions are shown in Fig. 3. For example, for an asteroid with a 5 hr rotation period observed at 1 AU, NESTM200 applies a night side temperature that is 0.58 × the maximum dayside temperature at zero degrees latitude. Figure 3 (c) shows the NEATM and NESTM equatorial temperatures.

Synthetic thermal IR fluxes $F_{obs}(n)$ were generated depending on the assigned parameters: asteroid diameter $D_{eff}$, Earth-asteroid distance $\Delta$ (AU), phase angle $\alpha$, "instrument" wavelengths $\lambda_{obs}(n)$. The $\lambda_{obs}$ were set at filter wavelengths equivalent to a range of narrow-band filters typically used for sampling a wide range of wavelengths: 4.8, 8.0, 8.9, 10.7, 11.7, 12.5 and 20.0 μm, i.e. one M- and Q-band measurement and five N-band measurements. The asteroid diameter $D_{eff}$ was set to 1.0 km, and $\Delta$ = 0.2 AU. The output flux is determined from the temperature array by integrating the blackbody function over the visible surface, i.e. over all latitudes, and for the 180° of longitude in the temperature distribution that would be visible depending on the phase angle:

$$F_{obs}(n) = \frac{\varepsilon D_{eff}^2}{4\Delta^2} \int_{-\frac{\pi}{2}}^{+\frac{\pi}{2}} \int_{\alpha-\frac{\pi}{2}}^{\alpha+\frac{\pi}{2}} B(\lambda_n, T(\theta,\phi)) \cos^2 \phi \cos(\alpha-\theta) \mathrm{d}\theta \, \mathrm{d}\phi \qquad (13)$$

The phase angle was varied for each asteroid and was set to: $\alpha$ = 0°, ±30°, ±45°, ±60°, ±75°, ±90°, ±105°, ±120°. The resulting thermal IR fluxes at 10.7 μm for each simulated surface are given in Fig. 4. The direction of the phase angle, i.e. whether the cooler morning side of the asteroid or the warmer afternoon side is being observed, is important. If we input negative $\alpha$ in Eq. 13 we can obtain a second set of results for the cooler morning side. This analysis assumes the extreme case of the pole orientation at 90° to the solar direction. In this geometry the effects of significant thermal inertia are at their greatest. If the spin axis is pointing towards the Sun, then no part of the day side is rotated onto the night side, there is no emission on the night side, and the NEATM is the appropriate model. In between, there is a gradation between the two cases.

The NEATM and NESTM were best-fitted to the thermal IR fluxes. $H_V$ was set to 17.12277 consistent with the test asteroid's 1 km diameter, following Eq. 2 (so we assume perfect precision in the optical observations). The derived effective diameters $D_{eff}$ are shown in Fig. 5. The $f$ parameters used for the different NESTM solutions varied only slightly from those given in the caption for Fig. 3, as the best-fit $p_v$ altered. The NEATM relative errors from the true diameter are consistent with the results of Delbó (2004), who performed a similar test to assess NEATM.

If we contrast the results with different surface thermal inertias observed between 45° and 75° phase angle (Fig. 6), we can see that on the afternoon side a range of different versions of NESTM produce more accurate diameters than NEATM. But on the morning side, if the simulated asteroid surface has low surface thermal inertia, then NEATM is more accurate. We find that NESTM200 produces the most improved accuracy of diameter estimation over the greatest range of asteroid surfaces. Note that Delbo *et al.* (2007) found that NEAs have an average surface thermal inertia of $200 \pm 40$ J m$^{-2}$ s$^{-1/2}$ K$^{-1}$ from a sample of size range 0.8 to 3.4 km diameter, and also that there is a trend of increasing thermal inertia with decreasing size. Therefore NESTM200 should be suitable for km-sized NEAs and smaller.

Figure 7 shows the variation of beaming parameter $\eta$ with phase angle $\alpha$ for NEATM and NESTM for the $\Gamma = 200$ J m$^{-2}$ s$^{-1/2}$ K$^{-1}$ simulated surface. The NEATM-derived $\eta$ are consistent with Delbó (2004). As the simulated surface increases in thermal inertia, so the $\eta$-value at zero phase angle increases. We have not included beaming (i.e. surface roughness) in our thermophysical model which would decrease $\eta$ at low phase angles and increase it at large phase angles. As the surface type increases in thermal inertia, the maximum day side temperature becomes reduced compared to $T_{max}$, conserving energy as more thermal flux comes from the night side. In the modified projected model this reduced maximum day side temperature $T_{mod}$ was calculated (Section 1.4) but in the NEATM and NESTM the observed temperature is effectively measured. As a result the best-fit $\eta$ increases.

The NESTM40, NESTM120 and NESTM200-derived $\eta$-values are relatively constant as a function of phase angle, therefore including appropriate thermal emission on the night side has the effect of reducing the increase of $\eta$. (If beaming were included in the model, the $\eta$-values would still increase to compensate for the enhanced thermal emission in the sunward direction, which would no longer be observed at larger phase angles.) The behaviour of the best-fit $\eta$ is quite complex. While the NEATM was inaccurate in not introducing night side thermal flux, NESTM forces a particular amount of thermal flux from the night side. The more night side flux is introduced the less the fitted maximum day side temperature $T_{fit}$ has to be reduced in order to account for the fact that the observed day side temperature is lower than $T_{max}$.

For high thermal inertia versions of NESTM fitted to low thermal inertia surfaces, very high best-fit $\eta$ can be derived at high phase angles. An extreme case is NESTM550 at $\alpha$ = -120°, which gives $\eta$ = 4.3. This behaviour does not seem to especially affect the accuracy of the fitted diameter ($\Delta D_{eff}$ = -19%) which is much closer to the true value than for NEATM ($\Delta D_{eff}$ = +117%). The reason for these strange results is that at such a high phase angle, there are many more surface elements from both the night side and the day side with temperatures replaced by $fT_{max} \cos^{1/4} \phi$, than there are elements fitted by $\eta$, and so the isothermal latitude FRM begins to fit the observed fluxes better.

### 3.2    Testing on Real Asteroid Spectra

The thermal IR fluxes of (33342) 1998 WT$_{24}$ reported by Harris *et al.* (2007) are a useful test of NESTM because the same asteroid is observed over a range of high phase angles (60°-93°). You would expect NEATM to increasingly overestimate the diameter, while NESTM diameters should be more consistent as the phase angle increases, which is the case (Fig. 8). Because (33342) 1998 WT$_{24}$ has been observed by radar (Zaitsev *et al.* 2002), constraints on its shape and pole orientation make it suitable for deriving a diameter with a thermophysical model. This was done by Harris *et al.*, who obtained $D_{eff}$ = 0.35 ± 0.04 km, $p_v$ = 0.56 ± 0.2 and $\Gamma$ = 100-300 J m$^{-2}$ s$^{-1/2}$ K$^{-1}$. As can be seen, NESTM agrees with this derived diameter better than NEATM, increasingly so at higher phase angles. We note that the FRM also fits well.

We can derive NESTM diameters for a number of objects whose diameters have been determined independently by radar. Delbo (2004) performed a similar study for NEATM using a dataset containing 10 objects (23 total spectra, 9 with default $\eta$) and found NEATM overestimated the diameter by +8% ± 4%. We performed the same analysis on the same data for NESTM and found NESTM200 gave 0% ± 4% (Wolters 2005).

Here we present a comparison using a wider dataset of 18 objects (62 spectra, 38 with default $\eta$). The results for NEATM, FRM and NESTM200 are shown in Tables 2-4. We have derived NESTM diameters using NESTM40, 120, 200, 550, and 2200 for this dataset. Figure 9 shows the relative diameter discrepancy ($D_{radiometry} - D_{radar}$)/$D_{radar}$ for each object, and the mean relative diameter discrepancy is given in Table 5. We have separated the dataset into observations where $\eta$ was fitted [Fig. 9 (a)] and when default $\eta$ was used [Fig. 9 (b)].

### 3.3    Discussion

When we use the subset of observations where $\eta$-fitting is used, then we can see that there is a systematic bias for both NEATM (+11% ± 8%) and the FRM (+18% ± 12%) which significantly overestimate the diameter in comparison to radar diameters, while NESTM200 (+3% ± 8%) and NESTM550 (-1% ± 8%) remove the bias within the uncertainty of the distribution of relative diameter errors.

However, we note that radar diameters themselves will have uncertainties which may be systematic (Ostro *et al.* 2002). In some cases, radar diameter uncertainty may be significantly larger than for thermal modelling. For example, for (1627) Ivar, the diameter estimation is 8.5 ± 3 km, a 35% uncertainty. It derives from a consideration in Ostro et al. (1990) over the uncertainty in sub-radar latitude $\delta$ (the angle between the radar line-of-sight and the asteroid's apparent equator), which alters the contribution of rotation phase to the range distribution of echo power. Problematically in a few cases, only a value for the lower bound on the maximum diameter ($D_{max}$) can be measured, e.g. $D_{max} > 0.46$ km for (3757) 1982 XB (Shepard et al., 2004). $D_{eff}$ could be smaller than this if the object is highly elongated, also it could be larger since it is a lower bound. In these cases we have assumed that $D_{eff}$ is equal to this value, and assigned an uncertainty of 15% (simply to match the uncertainty assigned to thermal model diameters). In a few cases, a very accurate effective diameter is obtained, for example $D_{eff} = 0.53 \pm 0.03$ km for (6489) Golevka (Hudson et al., 2000) (i.e. 6% uncertainty). In this case three-dimensional modelling using two-dimensional delay-Doppler images combined with published lightcurves could unambiguously define the pole and shape. Finally, in the cases of (433) Eros and (25143) Itokawa, we were able to use diameters measured by orbiting spacecraft, which can be considered a precise standard (1% and 2% uncertain respectively). Although the size of the radar uncertainty is influential when the sample size is small, more crucial for our comparison is whether there is a systematic bias (positive or negative) in the radar diameter estimation.

When we examine the results for the subset where default $\eta$ was used, then NEATM significantly underestimates the diameter (-16% ± 5%), while the discrepancy for NESTM is even greater. FRM diameters are closer to radar diameters on average than both NEATM and NESTM, though with a wider distribution (+7% ± 8%). The same default $\eta$ was used for NEATM and NESTM in this analysis; in some cases $\eta = 1.2$ whatever the phase angle (if this was the value used in the literature), in other cases $\eta = 1$ when $\alpha < 45°$ and $\eta = 1.5$ when $\alpha > 45°$, as suggested by Delbó *et al.* (2003). If the population of objects compared with radar diameters is assumed to be representative of the NEA population as a whole, then this suggests that the wrong default $\eta$ is being used for NEATM. The FRM makes no use of a beaming parameter which explains why the FRM is closer to the radar diameters.

There is a trend of increasing $\eta$ with phase angle; for a recent analysis see, e.g. Wolters *et al.* (2008) which found a trend $\eta = (0.013 \pm 0.004)\alpha + (0.91 \pm 0.17)$. However, there is a wide distribution, since $\eta$ is an observable that results from an amalgamation of influences: beaming, surface thermal inertia, pole orientation, shape etc. Nevertheless, future work may be able to improve NEATM diameters by using a more appropriate default $\eta$. We note that if a higher $\eta$ is used, the determined diameter is larger, so this analysis suggests that a higher default $\eta$ on average may be appropriate. The most appropriate default $\eta$ for NESTM will be different, since the model automatically supplies thermal IR flux from the night side. This has the consequence of lowering the best-fitted $\eta$ and flattening the trend with phase angle, as Fig. 10 illustrates. When the two datasets are combined, the systematic overestimation of diameter by NEATM when $\eta$ is fitted is compensated for by systematically underestimating the diameter when default $\eta$ is used, such that the overall distribution of NEATM diameters are not significantly different from radar.

## 4    Conclusions

NESTM has been introduced in an attempt to produce more reliable estimates of albedo and diameter of NEAs from thermal IR observations, when insufficient information is available for a thermophysical model to be applied. NESTM models the night side temperature as a fraction $f$ of the maximum day side temperature, with different versions applying different $f$ depending on the assumed thermal inertia. NEATM has zero night side emission and uses the beaming parameter $\eta$ as a free parameter to compensate. In NESTM, $\eta$ is also a free parameter, which can partly compensate for the modelled thermal inertia differing from an asteroid's actual thermal properties.

We find that NESTM does not remove phase effects sufficiently well to make NESTM derived $\eta$ more meaningful than those using NEATM. It is therefore not appropriate to use NESTM-derived best-fit $\eta$ for physical interpretation of asteroid surfaces, e.g. to estimate the true surface thermal inertia, as can be done for NEATM (e.g. Delbo, 2004). One of the merits of NEATM is its simplicity. The additional complexity of NESTM is only justified if it produces significantly improved accuracy for diameter determination.

NESTM200 (i.e. with assumed surface thermal inertia $\Gamma = 200 \pm 40$ J m$^{-2}$ s$^{-1/2}$ K$^{-1}$) produces significantly improved accuracy for diameter estimation only for observations taken at $\alpha > 45°$, and in some circumstances can be less accurate than NEATM for $\alpha < 45°$. In cases where there is not enough physical information about the asteroid or high enough quality data to use a thermophysical model, we suggest adopting NEATM for asteroids observed at $\alpha < 45°$ and NESTM200 as the default simple thermal model to apply to NEAs observed at $\alpha > 45°$.

A comparison of radar diameters derived from radar observations with those calculated using NEATM with $\eta$-fitting showed that NEATM significantly overestimates the diameter (+11% ± 8%), a bias which was removed within the uncertainties by using NESTM200 (+3% ± 8%). This may be due to a systematic bias in the diameter estimation of NEATM, and/or systematic bias in radar diameters. When default $\eta$ was used, the diameter was significantly underestimated by NEATM (-16% ± 5%). This suggests that the default $\eta$ for NEATM needs to be re-investigated.

**Acknowedgements**

We are grateful to the reviewer Dr. A. W. Harris for several suggested refinements to the manuscript that improved it noticeably. The work of S.D. Wolters is supported by the UK Science and Technology Facilities Council (STFC).

**Tables**

Table 1 : Look-up Table for *f*-parameter

| Thermal parameter Θ | *f* parameter | Thermal parameter Θ | *f* parameter |
|---|---|---|---|
| 0.058 | 0.326 | 1.133 | 0.584 |
| 0.065 | 0.334 | 1.266 | 0.594 |
| 0.075 | 0.345 | 1.462 | 0.607 |
| 0.082 | 0.351 | 1.602 | 0.615 |
| 0.092 | 0.360 | 1.791 | 0.624 |
| 0.106 | 0.371 | 2.068 | 0.636 |
| 0.116 | 0.379 | 2.312 | 0.644 |
| 0.130 | 0.388 | 2.532 | 0.651 |
| 0.150 | 0.400 | 2.831 | 0.659 |
| 0.168 | 0.409 | 3.203 | 0.668 |
| 0.184 | 0.417 | 3.269 | 0.669 |
| 0.206 | 0.426 | 3.581 | 0.675 |
| 0.238 | 0.439 | 4.004 | 0.681 |
| 0.260 | 0.447 | 4.135 | 0.683 |
| 0.291 | 0.457 | 4.530 | 0.688 |
| 0.304 | 0.461 | 4.623 | 0.689 |
| 0.318 | 0.465 | 5.065 | 0.694 |
| 0.336 | 0.470 | 5.663 | 0.699 |
| 0.353 | 0.475 | 5.848 | 0.700 |
| 0.376 | 0.481 | 6.406 | 0.704 |
| 0.412 | 0.489 | 7.163 | 0.708 |
| 0.439 | 0.495 | 8.008 | 0.712 |
| 0.460 | 0.500 | 8.271 | 0.713 |
| 0.485 | 0.505 | 9.247 | 0.716 |
| 0.515 | 0.510 | 10.129 | 0.719 |
| 0.550 | 0.516 | 11.325 | 0.722 |
| 0.582 | 0.522 | 13.077 | 0.725 |
| 0.594 | 0.524 | 14.325 | 0.726 |
| 0.651 | 0.532 | 16.016 | 0.728 |
| 0.728 | 0.543 | 18.494 | 0.730 |
| 0.801 | 0.552 | 22.650 | 0.732 |
| 0.895 | 0.562 | 32.032 | 0.733 |
| 1.034 | 0.576 | | |

Table 2: Original data sources for NEAs for which radar diameters and thermal IR fluxes are available

| Asteroid | date yyyy-mm-dd | Thermal IR flux source | NEATM fit source | $H_V$ mag. source | Radar diameter source |
|---|---|---|---|---|---|
| (433) Eros | 1975-01-17 | Lebofsky and Rieke (1979) | Harris (1998) | Harris (1998) – lc max | Harris and Lagerros (2002) [a] |
| | 1998-06-27 | Harris and Davies (1999) | Harris and Davies (1999) | Harris and Davies (1999) - lc max | |
| | 2002-09-21, -22 | Lim *et al.* (2005) | Wolters *et al.* (2008) | Wolters *et al.* (2008) | |
| | 2002-09-28 | Wolters *et al.* (2008) | Wolters *et al.* (2008) | Wolters *et al.* (2008) | |
| (1566) Icarus | 1987-06-22, -23 | Veeder *et al.* (1989) | This work | Harris (1998) | Goldstein (1968) |
| | 1987-06-23.32 | Veeder *et al.* (1989) | Harris (1998) | Harris (1998) | |
| (1580) Betulia | 1976-05-17 | Lebofsky *et al.* (1978) | Harris (1998) | Harris (1998) | Magri *et al.* (2007) |
| | 2002-06-02 | Harris *et al.* (2005) | Harris *et al.* (2005) | Harris *et al.* (2005) | |
| (1620) Geographos | 1983-11-03.31 | Veeder *et al.* (1989) | Harris (1998) | Harris (1998) - lc max | Hudson and Ostro (1999) |
| (1627) Ivar | 1980-11-26.42 | Veeder *et al.* (1989) | This work | Hahn *et al.* (1998) | Ostro *et al.* (1990) |
| | 1985-07-10 | Veeder *et al.* (1989) | Harris (1998) | Harris (1998) - lc max | |
| | 2000-03-06 | Delbó *et al.* (2003) [b] | Delbó *et al.* (2003) | Delbó *et al.* (2003) | |
| (1685) Toro | 1981-03-12 | Veeder *et al.* (1989) | Harris (1998) | Harris (1998) | Ostro *et al.* (1983) |
| (1862) Apollo | 1980-11-26 | Lebofsky *et al.* (1981) [c] | Harris (1998) | Harris (1998) | Goldstein *et al.* (1981) |
| (1915) Quetzalcoatl | 1981-03-12 | Veeder *et al.* (1989) | Harris (1998) | Harris (1998) | Shepard *et al.* (2004) |
| (2100) Ra-Shalom | 1978-09-13.17 | Lebofsky *et al.* (1979) | This work | Pravec *et al.* (1998) | Ostro *et al.* (1984) |
| | 1981-08-22 → 1981-08-24 | Veeder *et al.* (1989) | This work | Pravec *et al.* (1998) | |
| | 1997-08-31 | Harris *et al.* (1998) | Harris *et al.* (1998) | Harris *et al.* (1998) | |
| | 2000-08-21 | Delbó *et al.* (2003) [b] | Delbó *et al.* (2003) | Delbó *et al.* (2003) | |
| (3103) Eger | 1986-07-02 → 1987-01-27 | Veeder *et al.* (1989) | This work | Pravec *et al.* (1998) | Benner *et al.* (1997) |
| (3757) 1982 XB | 1982-12-16 | Veeder *et al.* (1989) | Harris and Lagerros (2002) | Minor Planet Center | Shepard *et al.* (2004) |
| | 1982-12-17 | Veeder *et al.* (1989) | This work | Minor Planet Center | |
| (3908) Nyx | 1988-09-08 | Cruikshank *et al.* (1991) | This work | Cruikshank *et al.* (1991) | Benner *et al.* (2002) |
| (5381) Sekhmet | 2003-05-12 → 2003-05-15 | Delbo (2004) | Delbo (2004) | Minor Planet Center | Nolan *et al.* (2003) |
| | 2003-05-13, -05-16, -06-02 | Delbo (2004) | This work | Minor Planet Center | |
| (6178) 1986 DA | 1986-03-13, -05-22 | Tedesco and Gradie (1987) | This work [d] | Ondrejov Asteroid Photometry Project [e] | Ostro *et al.* (1991) |
| (6489) Golevka | 1995-06-16 | Mottola *et al.* (1997) | Harris (1998) | Harris (1998) | Hudson *et al.* (2000) |
| | 2003-05-15 | Delbo (2004) | This work | Mottola *et al.* (1997) | |
| (25143) Itokawa | 2001-03-14 | Sekiguchi *et al.* (2003) [f] | This work | Sekiguchi *et al.* (2003) | Fujiwara *et al.* (2006) [a] |
| | 2001-04-08, -09 | Delbo (2004) [f] | Delbo (2004) | Sekiguchi *et al.* (2003) | |
| | 2004-07-01 | Müller *et al.* (2005) | This work | Sekiguchi *et al.* (2003) | |
| | 2004-07-10 | Mueller *et al.* (2007) | This work | Sekiguchi *et al.* (2003) | |
| (33342) 1998 WT24 | 2001-12-04, -19, -21 | Harris *et al.* (2007) | Harris *et al.* (2007) | Harris *et al.* (2007) | Zaitsev *et al.* (2002) |
| | 2001-12-18 | Harris *et al.* (2007) | This work | Harris *et al.* (2007) | |
| 2002 BM26 | 2002-02-21 | Delbó *et al.* (2003) | Delbó *et al.* (2003) | Minor Planet Center | Nolan *et al.* (2002) |

Notes

[a] Used spacecraft diameter.

[b] Lightcurve corrected fluxes (Delbo, private communication, 2004)

[c] Lightcurve corrected as described in Harris (1998)

[d] Harris and Lagerros (2002) may have used average of two magnitudes for 1986-05-22 data, we derive fits for both. They measured $p_v = 0.17$, $D_{eff} = 2.1$ km

[e] P. Pravec and colleagues. http://www.asu.cas.cz/~ppravec/neo.html

[f] Used re-calibrated fluxes from Müller *et al.* (2005). Conditions on 2001-04-09 were "less favourable".

Table 3: Thermal IR observational circumstances of asteroids observed by both thermal IR and radar

| Asteroid | Type | P (h) | date yyyy-mm-dd | $H_V$ | G | r (AU) | Δ (AU) | α (°) |
|---|---|---|---|---|---|---|---|---|
| (433) Eros | S | 5.270 | 1975-01-17 | 10.47 | 0.15 | 1.134 | 0.153 | 9.9 |
| | | | 1998-06-27 | 10.47 | 0.32 | 1.619 | 0.804 | 30.9 |
| | | | 2002-09-21.25 | 10.36 | 0.15 | 1.608 | 0.637 | 14.5 |
| | | | 2002-09-22.23 | 10.39 | | 1.606 | 0.637 | 15.0 |
| | | | 2002-09-22.27 | 10.34 | | 1.606 | 0.637 | 15.0 |
| | | | 2002-09-28 | 10.32 | 0.15 | 1.589 | 0.640 | 18.2 |
| (1566) Icarus | Q | 2.273 | 1987-06-22.26 | 16.3 | 0.09 | 0.9771 | 0.1624 | 99.4 |
| | | | 1987-06-22.28 | | | 0.9773 | 0.1625 | 99.3 |
| | | | 1987-06-22.32 | | | 0.9779 | 0.1626 | 99.1 |
| | | | 1987-06-23.29 | | | 0.9924 | 0.1665 | 93.6 |
| | | | 1987-06-23.32 | | | 0.9929 | 0.1666 | 93.4 |
| (1580) Betulia | C | 6.138 | 1976-05-23 | 14.58 | 0.18 | 1.140 | 0.130 | 10 |
| | | | 2002-06-02 | 15.1 | 0.15 | 1.143 | 0.246 | 53 |
| (1620) Geographos | S | 5.223 | 1983-11-03.31 | 15.09 | 0.31 | 1.070 | 0.095 | 34 |
| (1627) Ivar | S | 4.795 | 1980-11-26.42 | 13.24 | 0.25 | 1.834 | 0.919 | 16 |
| | | | 1985-07-10 | 12.9 | 0.25 | 1.124 | 0.202 | 53 |
| | | | 2000-03-06 | 12.87 | 0.25 | 2.057 | 1.073 | 5 |
| (1685) Toro | S | 10.196 | 1981-03-12 | 13.9 | 0.07 | 1.668 | 0.738 | 18 |
| (1862) Apollo | Q | 3.065 | 1980-11-26 | 16.27 | 0.23 | 1.105 | 0.148 | 35 |
| (1915) Quetzalcoatl | S | 4.9 | 1981-03-12 | 18.9 | 0.06 | 1.095 | 0.117 | 29 |
| (2100) Ra-Shalom | C | 19.800 | 1978-09-13.17 | 16.07 | 0.12 | 1.194 | 0.189 | 4 |
| | | | 1981-08-22.45 | | | 1.155 | 0.180 | 34 |
| | | | 1981-08-24.32 | | | 1.158 | 0.180 | 33 |
| | | | 1997-08-31 | 15.9 | 0.12 | 1.195 | 0.264 | 40.6 |
| | | | 2000-08-21 | 16.11 | 0.12 | 1.175 | 0.222 | 39 |
| (3103) Eger | E | 5.7059 | 1986-07-02.48 | 15.74 | 0.4 | 1.259 | 0.384 | 44 |
| | | | 1986-07-02.52 | | | 1.259 | 0.384 | 44 |
| | | | 1986-07-03.50 | | | 1.253 | 0.377 | 44 |
| | | | 1986-07-03.53 | | | 1.253 | 0.377 | 44 |
| | | | 1986-07-03.56 | | | 1.253 | 0.377 | 44 |
| | | | 1986-08-03.64 | | | 1.071 | 0.145 | 64 |
| | | | 1986-08-04.52 | | | 1.066 | 0.144 | 65 |
| | | | 1986-08-04.54 | | | 1.066 | 0.144 | 65 |
| | | | 1986-08-04.56 | | | 1.066 | 0.144 | 65 |
| | | | 1986-08-04.58 | | | 1.066 | 0.144 | 65 |
| | | | 1987-01-25.45 | | | 1.415 | 0.478 | 21 |
| | | | 1987-01-27.47 | | | 1.420 | 0.479 | 19 |
| (3757) 1982 XB | S | 9.0046 | 1982-12-16 | 18.95 | 0.25 | 1.021 | 0.043 | 32 |
| | | | 1982-12-17 | | | 1.019 | 0.041 | 30[a] |
| (3908) Nyx | S | 4.426 | 1988-09-08 | 17.56 | 0.4 | 1.135 | 0.135 | 17.9 |
| (5381) Sekhmet | V | 3.6 | 2003-05-12 | 16.5 | 0.4 | 1.114 | 0.146 | 42 |
| | | | 2003-05-13 | | | 1.117 | 0.140 | 38 |
| | | | 2003-05-14 | | | 1.121 | 0.135 | 33 |
| | | | 2003-05-15 | | | 1.124 | 0.132 | 29 |
| | | | 2003-05-16 | | | 1.228 | 0.129 | 24 |
| | | | 2003-06-02 | | | 1.176 | 0.247 | 44 |
| (6178) 1986 DA | M | 3.58 | 1986-03-13 | 15.94 | 0.25 | 1.205 | 0.246 | 27.6 |
| | | | 1986-05-22.26 | | | 1.179 | 0.211 | 31.4 |
| | | | 1986-05-22.48 | | | 1.180 | 0.211 | 31.4 |
| (6489) Golevka | Sq | 6.026 | 1995-06-16 | 18.82 | 0.14 | 1.016 | 0.051 | 88.8 |
| | | | 2003-05-15 | 19.07 | 0.14 | 1.081 | 0.099 | 43 |
| (25143) Itokawa | S(IV) | 12.1324 | 2001-03-14 | 19.48 | 0.21 | 1.0592 | 0.074 | 27.5 |
| | | | 2001-04-08 | 19.9 | 0.21 | 0.983 | 0.054 | 108 |
| | | | 2001-04-09 | | | 0.981 | 0.056 | 110 |
| | | | 2004-07-01 | | | 1.028 | 0.020 | 54 |
| | | | 2004-07-10 | | | 1.061 | 0.050 | 29 |
| (33342) 1998 WT24 | E | 3.723 | 2001-12-04 | 18.5 | 0.4 | 1.0148 | 0.0621 | 60.4[b] |
| | | | 2001-12-18 | | | 0.9901 | 0.0162 | 67.5 |
| | | | 2001-12-19 | | | 0.9874 | 0.0198 | 79.3 |
| | | | 2001-12-21 | | | 0.9817 | 0.0284 | 93.4 |
| 2002 BM26 | P | 2.7 | 2002-02-21 | 20.1 | 0.15 | 1.024 | 0.074 | 60 |

Notes

[a] $r$, $\Delta$, $\alpha$ in Veeder *et al.* (1989) are incorrect for given time of observation and give much higher albedo inconsistent with their other data

[b] According to Harris *et al.* (2007) the sense of the solar phase angle changed on December 15. So 12-04-2001 is morning side, and the rest are afternoon, or visa versa.

Table 4
NEATM, FRM and NESTM200 derived $p_v$, $D_{eff}$, and $\eta$, $f$ parameters used, and radar diameters for NEAs for which both radar diameters and thermal IR fluxes are available

| Asteroid | Date yyyy-mm-dd | α (°) | Radar D (km) | ±a (km) | FRM $p_v$ | FRM $D_{eff}$ (km) | NEATM $p_v$ | NEATM $D_{eff}$ (km) | NEATM $\eta^b$ | NESTM200 $p_v{}^c$ | NESTM200 $D_{eff}$ (km) | NESTM200 $\eta$ | f |
|---|---|---|---|---|---|---|---|---|---|---|---|---|---|
| | 1975-01-17 | 9.9 | | | 0.09 | 35.88 | 0.20 | 23.6 | 1.05 | 0.21 | 23.36 | 1.06 | 0.61 |
| | 1998-06-27 | 30.9 | | | 0.08 | 37.84 | 0.21 | 23.6 | 1.07 | 0.22 | 22.82 | 1.00 | 0.64 |
| | 2002-09-21.25 | 14.5 | | | 0.09 | 37.53 | 0.32 | 19.9 | 0.80 | 0.32 | 19.84 | 0.80 | 0.64 |
| | 2002-09-22.23 | 15.0 | | | 0.09 | 37.02 | 0.29 | 20.7 | 0.83 | 0.29 | 20.55 | 0.84 | 0.64 |
| | 2002-09-22.27 | 15.0 | | | 0.10 | 35.94 | 0.32 | 20.0 | 0.79 | 0.33 | 19.84 | 0.78 | 0.64 |
| | *average* | 14.8 | | | 0.09 | 36.8 | 0.31 | 20.20 | 0.81 | 0.31 | 20.1 | 0.80 | |
| | 2002-09-28 | 18.2 | | | 0.08 | 40.30 | 0.24 | 23.31 | 0.95 | 0.25 | 23.12 | 0.94 | 0.64 |
| **(433) Eros** | average | **18.5** | **20.06** | **0.2** | **0.09** | **37.71** | **0.24** | **22.7** | **0.97** | **0.11** | **22.34** | **0.95** | |
| | 1987-06-22.26 | 99.4 | | | 0.68 | 0.89 | 0.41 | 1.14 | (1.2) | 0.52 | 1.02 | (1.2) | 0.62 |
| | 1987-06-22.28 | 99.3 | | | 0.80 | 0.82 | 0.50 | 1.04 | (1.2) | 0.65 | 0.91 | (1.2) | 0.62 |
| | 1987-06-22.32 | 99.1 | | | 0.71 | 0.87 | 0.43 | 1.12 | (1.2) | 0.57 | 0.97 | (1.2) | 0.62 |
| | 1987-06-23.29 | 93.6 | | | 0.49 | 1.04 | 0.33 | 1.27 | (1.2) | 0.42 | 1.12 | (1.2) | 0.62 |
| | 1987-06-23.32 | 93.4 | | | 0.49 | 1.04 | 0.33 | 1.27 | (1.2) | 0.42 | 1.12 | (1.2) | 0.62 |
| **(1566) Icarus** | average | **96.4** | **1.0** | | **0.63** | **0.93** | **0.40** | **1.17** | | **0.52** | **1.03** | | |
| | 1976-05-23 | 10 | | | 0.08 | 5.70 | 0.17 | 3.9 | (1.2) | 0.17 | 3.96 | (1.2) | 0.59 |
| | 2002-06-02 | 53 | | | 0.07 | 4.80 | 0.11 | 3.82 | 1.09 | 0.13 | 3.59 | 1.00 | 0.59 |
| **(1580) Betulia** | average | **31.5** | **5.39** | | **0.08** | **5.25** | **0.14** | **3.86** | | **0.15** | **3.77** | | |
| **(1620) Geographos** | 1983-11-03.31 | **34** | **2.56** | | **0.26** | **2.50** | **0.26** | **2.5** | (1.2) | **0.26** | **2.49** | (1.2) | 0.59 |
| | 1980-11-26.42 | 16 | | | 0.08 | 10.57 | 0.26 | 5.92 | (1.0) | 0.26 | 5.91 | (1.0) | 0.67 |
| | 1985-07-10 | 53 | | | 0.08 | 12.36 | 0.12 | 10.2 | (1.2) | 0.12 | 9.97 | (1.2) | 0.61 |
| | 2000-03-06 | 5 | | | 0.05 | 15.85 | 0.15 | 9.12 | (1.0) | 0.16 | 8.94 | (1.0) | 0.68 |
| **(1627) Ivar** | average | **24.7** | **8.5** | **3** | **0.07** | **12.93** | **0.18** | **8.41** | | **0.18** | **8.27** | | |
| **(1685) Toro** | 1981-03-12 | **18** | **3.3** | **0.9** | **0.12** | **6.37** | **0.29** | **4.1** | (1.2) | **0.29** | **4.07** | (1.2) | 0.62 |
| **(1862) Apollo** | 1980-11-26 | **35** | **1.2** | | **0.16** | **1.85** | **0.26** | **1.45** | 1.15 | **0.29** | **1.38** | 1.07 | 0.62 |
| **(1915) Quetzalcoatl** | 1981-03-12 | **29** | **0.75** | **0.25** | **0.16** | **0.55** | **0.31** | **0.40** | (1.2) | **0.30** | **0.40** | (1.2) | 0.61 |
| | 1978-09-13.17 | 4 | | | 0.07 | 3.07 | 0.18 | 1.94 | (1.0) | 0.18 | 1.94 | (1.0) | 0.54 |
| | 1981-08-22.45 | 34 | | | 0.08 | 2.87 | 0.18 | 1.94 | (1.0) | 0.18 | 1.93 | (1.0) | 0.54 |
| | 1981-08-24.32 | 33 | | | 0.09 | 2.71 | 0.22 | 1.74 | (1.0) | 0.22 | 1.73 | (1.0) | 0.54 |
| | *average* | 34 | | | 0.09 | 2.79 | 0.20 | 1.84 | | 0.20 | 1.83 | | |
| | 1997-08-31 | 40.6 | | | 0.11 | 2.65 | 0.13 | 2.48 | 1.80 | 0.14 | 2.36 | 1.69 | 0.55 |
| | 2000-08-21 | 39 | | | 0.10 | 2.52 | 0.08 | 2.79 | 2.32 | 0.09 | 2.64 | 2.22 | 0.55 |
| **(2100) Ra-Shalom** | average | **29.3** | **2.4** | | **0.09** | **2.76** | **0.15** | **2.26** | | **0.15** | **2.19** | | |
| | 1986-07-02.48 | 44 | | | 0.38 | 1.53 | 0.68 | 1.15 | (1.0) | 0.69 | 1.14 | (1.0) | 0.62 |
| | 1986-07-02.52 | 44 | | | 0.55 | 1.28 | 0.90 | 0.99 | (1.0) | 0.91 | 0.99 | (1.0) | 0.65 |
| | *average* | 44 | | | 0.47 | 1.41 | 0.79 | 1.07 | | 0.80 | 1.07 | | |
| | 1986-07-03.50 | 44 | | | 0.31 | 1.70 | 0.58 | 1.24 | (1.0) | 0.59 | 1.23 | (1.0) | 0.64 |
| | 1986-07-03.53 | 44 | | | 0.33 | 1.65 | 0.62 | 1.21 | (1.0) | 0.62 | 1.20 | (1.0) | 0.64 |
| | 1986-07-03.56 | 44 | | | 0.44 | 1.43 | 0.76 | 1.08 | (1.0) | 0.77 | 1.08 | (1.0) | 0.64 |
| | *average* | 44 | | | 0.36 | 1.59 | 0.65 | 1.18 | | 0.66 | 1.17 | | |
| | 1986-08-03.64 | 64 | | | 0.49 | 1.35 | 0.68 | 1.15 | (1.0) | 0.70 | 1.13 | (1.0) | 0.62 |
| | 1986-08-04.52 | 65 | | | 0.55 | 1.28 | 0.74 | 1.10 | (1.0) | 0.76 | 1.08 | (1.0) | 0.62 |
| | 1986-08-04.54 | 65 | | | 0.51 | 1.32 | 0.69 | 1.14 | (1.0) | 0.71 | 1.12 | (1.0) | 0.62 |
| | 1986-08-04.56 | 65 | | | 0.57 | 1.25 | 0.77 | 1.08 | (1.0) | 0.79 | 1.06 | (1.0) | 0.62 |
| | 1986-08-04.58 | 65 | | | 0.49 | 1.35 | 0.67 | 1.15 | (1.0) | 0.70 | 1.13 | (1.0) | 0.62 |
| | *average* | 65 | | | 0.53 | 1.30 | 0.72 | 1.12 | | 0.74 | 1.10 | | |
| | 1987-01-25.45 | 21 | | | 0.23 | 1.97 | 0.54 | 1.28 | (1.0) | 0.55 | 1.28 | (1.0) | 0.64 |
| | 1987-01-27.47 | 19 | | | 0.25 | 1.89 | 0.58 | 1.24 | (1.0) | 0.58 | 1.24 | (1.0) | 0.64 |
| **(3103) Eger** | average | **43** | **1.7** | | **0.39** | **1.58** | **0.66** | **1.17** | | **0.67** | **1.16** | | |
| | 1982-12-16 | 32 | | | 0.18 | 0.51 | 0.34 | 0.39 | (1.2) | 0.33 | 0.37 | (1.2) | 0.56 |
| | 1982-12-17 | 30 | | | 0.17 | 0.52 | 0.31 | 0.39 | (1.2) | 0.31 | 0.39 | (1.2) | 0.56 |
| **(3757) 1982 XB** | average | **31** | **0.46** | | **0.18** | **0.52** | **0.33** | **0.39** | | **0.32** | **0.38** | | |
| **(3908) Nyx** | 1988-09-08 | **17.9** | **1.04** | **0.16** | **0.28** | **0.77** | **0.57** | **0.54** | (1.0) | **0.58** | **0.54** | (1.0) | 0.62 |
| | 2003-05-12 | 42 | | | 0.19 | 1.53 | 0.25 | 1.34 | 1.52 | 0.26 | 1.3 | 1.44 | 0.62 |
| | 2003-05-13 | 38 | | | 0.18 | 1.57 | 0.16 | 1.67[d] | 2.34 | 0.17 | 1.63 | 2.27 | 0.62 |
| | 2003-05-14 | 33 | | | 0.18 | 1.57 | 0.30 | 1.21 | 1.22 | 0.32 | 1.18 | 1.18 | 0.62 |
| | 2003-05-15 | 29 | | | 0.17 | 1.62 | 0.23 | 1.38 | 1.59 | 0.24 | 1.35 | 1.55 | 0.62 |
| | 2003-05-16 | 24 | | | 0.13 | 1.85 | 0.25 | 1.34[e] | 1.22 | 0.26 | 1.32 | 1.19 | 0.64 |
| | *average* | 33 | | | 0.17 | 1.63 | 0.24 | 1.39 | | 0.25 | 1.36 | | |
| | 2003-06-02 | 44 | | | 0.18 | 1.57 | 0.16 | 1.65[f] | 2.10 | 0.18 | 1.59 | 2.00 | 0.62 |
| **(5381) Sekhmet** | average | **39** | **1.04** | | **0.18** | **1.60** | **0.20** | **1.52** | | **0.21** | **1.47** | | |
| | 1986-03-13 | 27.6 | | | 0.05 | 3.86 | 0.09 | 2.87 | | 0.09 | 2.86 | (1.0) | 0.62 |
| | 1986-05-22.26 | 31.4 | | | 0.07 | 3.26 | 0.15 | 2.23 | (1.0) | 0.15 | 2.21 | (1.0) | 0.62 |
| | 1986-05-22.48 | 31.4 | | | 0.08 | 3.05 | 0.18 | 2.03[g] | (1.0) | 0.18 | 2.02 | (1.0) | 0.62 |
| | *average* | 31.4 | | | 0.08 | 3.16 | 0.17 | 2.13 | | 0.17 | 2.12 | | |
| **(6178) 1986 DA** | average | **29.5** | **2.3** | **0.6** | **0.06** | **3.51** | **0.13** | **2.50** | | **0.13** | **2.49** | | |
| | 1995-06-16 | 88.8 | | | 0.79 | 0.26 | 0.63 | 0.29 | (1.2) | 0.71 | 0.27 | (1.2) | 0.59 |
| | 2003-05-15 | 43 | | | 0.19 | 0.47 | 0.24 | 0.42 | 1.61 | 0.25 | 0.41 | 1.55 | 0.59 |
| **(6489) Golevka** | average | **65.9** | **0.53** | **0.03** | **0.49** | **0.37** | **0.44** | **0.35** | | **0.48** | **0.34** | | |
| | 2001-03-14 | 27.5 | | | 0.19 | 0.39 | 0.38 | 0.28[h] | (1.0) | 0.38 | 0.27 | (1.0) | 0.56 |
| | 2001-04-08 | 108 | | | 0.41 | 0.22 | 0.13 | 0.38[i] | 1.54 | 0.28 | 0.26 | 0.91 | 0.55 |
| | 2001-04-09 | 110 | | | 0.33 | 0.24 | 0.10 | 0.45 | (1.5) | 0.12 | 0.40 | (1.5) | 0.54 |
| | *average* | 109 | | | 0.37 | 0.23 | 0.12 | 0.41 | | 0.20 | 0.33 | | |
| | 2004-07-01 | 54 | | | 0.22 | 0.30 | 0.24 | 0.28[j] | (1.5)[k] | 0.25 | 0.28 | (1.5) | 0.55 |
| | 2004-07-10 | 29 | | | 0.10 | 0.44 | 0.22 | 0.30[l] | 0.89 | 0.23 | 0.29 | 0.87 | 0.55 |
| **(25143) Itokawa** | average | **54.9** | **0.328** | **0.006** | **0.22** | **0.34** | **0.24** | **0.32** | | **0.26** | **0.30** | | |
| | 2001-12-04 | 60.4 | | | 0.50 | 0.38 | 0.44 | 0.40 | 1.86 | 0.51 | 0.37 | 1.62 | 0.62 |
| | 2001-12-18 | 67.5 | | | 0.47 | 0.39 | 0.42 | 0.41[m] | (1.5) | 0.45 | 0.40 | (1.5) | 0.62 |
| | 2001-12-19 | 79.3 | | | 0.49 | 0.38 | 0.42 | 0.41 | 1.26 | 0.59 | 0.35 | 0.92 | 0.62 |
| | 2001-12-21 | 93.4 | | | 0.57 | 0.35 | 0.19 | 0.61[n] | 2.67 | 0.34 | 0.45 | 1.72 | 0.61 |
| **(33342) 1998 WT24** | average | **75.2** | **0.41** | | **0.51** | **0.38** | **0.37** | **0.46** | | **0.47** | **0.39** | | |
| **2002 BM26** | 2002-02-21 | **60** | **0.61** | | **0.050** | **0.57** | **0.02** | **0.84** | 3.07 | **0.03** | **0.77** | 2.78 | 0.62 |

Notes.

[a] If uncertainty is not given in literature or only a lower bound of $D_{max}$ is given, uncertainty is assumed to be 15%.

[b] $\eta$-values in brackets are the default $\eta$ used and are not best-fitted, due to their only being one or two flux values, or for other reasons explained in the NEATM fit source (Table 2). Uncertainty is estimated as 20% (e.g. Delbó et al. 2003).

[c] As is commonly assumed for the NEATM (e.g. Delbó et al. 2003), the uncertainty in $p_v$, $D_{eff}$ and $\eta$ is assumed to be 30%, 15% and 20% respectively for both NEATM and NESTM when calculating the formal uncertainty of the relative diameter error $\sigma_{rel\_D}$.

[d] Delbo (2004) reported $p_v$ =0.24, $D_{eff}$ = 1.4 km, $\eta$=1.7 apparently fitting the same data with NEATM. It is possible his data was actually lc-corrected.

[e] Delbo (2004) reported $p_v$ = 0.22, $D_{eff}$ = 1.4 km.

[f] Delbo (2004) reported $D_{eff}$ = 1.5 km and $p_v$ = 0.22 from lc-corrected fluxes, and $D_{eff}$ = 1.65 km and $p_v$ = 0.16 from uncorrected fluxes. We have fitted lc-corrected fluxes and acquire a very similar $p_v$ and $D_{eff}$ to uncorrected fluxes, but with slightly lower $\eta$.

[g] Harris and Lagerros (2002) may have used average of two magnitudes for 1986-05-22 data and measured $p_v$ = 0.17, $D_{eff}$ = 2.1 km.

[h] Sekiguchi et al. (2003) found $p_v$ = 0.23, $D_{eff}$ = 0.35 km using a modified STM.

[i] Delbo (2004) obtained $p_v$ =0.19, $D_{eff}$ = 0.37 km using default $\eta$ = 1.5. Discrepancy may result if different $H_V$ was assumed for this date.

[j] Müller *et al.* (2005) obtained $p_v$ = 0.19 (+0.11, -0.03), $D_{eff}$ =0.32 ± 0.03 km using a thermophysical model (TPM) on combined Delbo (2004) + Sekiguchi *et al.* (2003) + their data.

[k] Using default $\eta$ because fitted $\eta$ = 0.33 (giving $p_v$ = 0.76). Probably due to rotation giving different surface area over 3 hours of observations, since these are lightcurve-uncorrected thermal fluxes (which is appropriate for TPM).

[l] Mueller obtained diameter of 0.28 km using TPM combining 2001-07-10 data with 2001-03-14 data (Sekiguchi et al. 2003), 2001-04-08, 2001-04-09 data and a 4.68 um observation in Ishiguro *et al.* (2003).

[m] Harris *et al.* (2007) obtain $p_v$ = 0.75, $D_{eff}$ = 0.31, $\eta$ = 0.61, but suggest that data may be of poor quality. We re-derive using default $\eta$ = 1.5.

[n] Harris et al. (2007) also report a TPM result combining all NASA IRTF data and obtain $p_v$ = 0.56 ± 0.2, $D_{eff}$ = 0.35 ± 0.04 km, thermal inertia $\Gamma$ = 200 ± 100 J m$^{-2}$ s$^{-0.5}$ K$^{-1}$.

Table 5 Mean relative discrepancy between radiometric and radar diameters

| Model | Mean Relative Discrepancy (%) | | |
|---|---|---|---|
| | Fitted η only | Default η only | All Observations |
| NEATM | +11 | -16 | -2 |
| FRM | +18 (±12) | +7 (±8) | +17 (±9) |
| NESTM40 | +9 | -16 | -3 |
| NESTM120 | +5 | -17 | -5 |
| NESTM200 | +3 | -18 | -6 |
| NESTM550 | -1 | -19 | -8 |
| NESTM2200 | -3 | -20 | -10 |
| *Standard error of NEATM/NESTM distributions* | ±8 | ±5 | ±6 |

**Figures**
*Fig. 1*

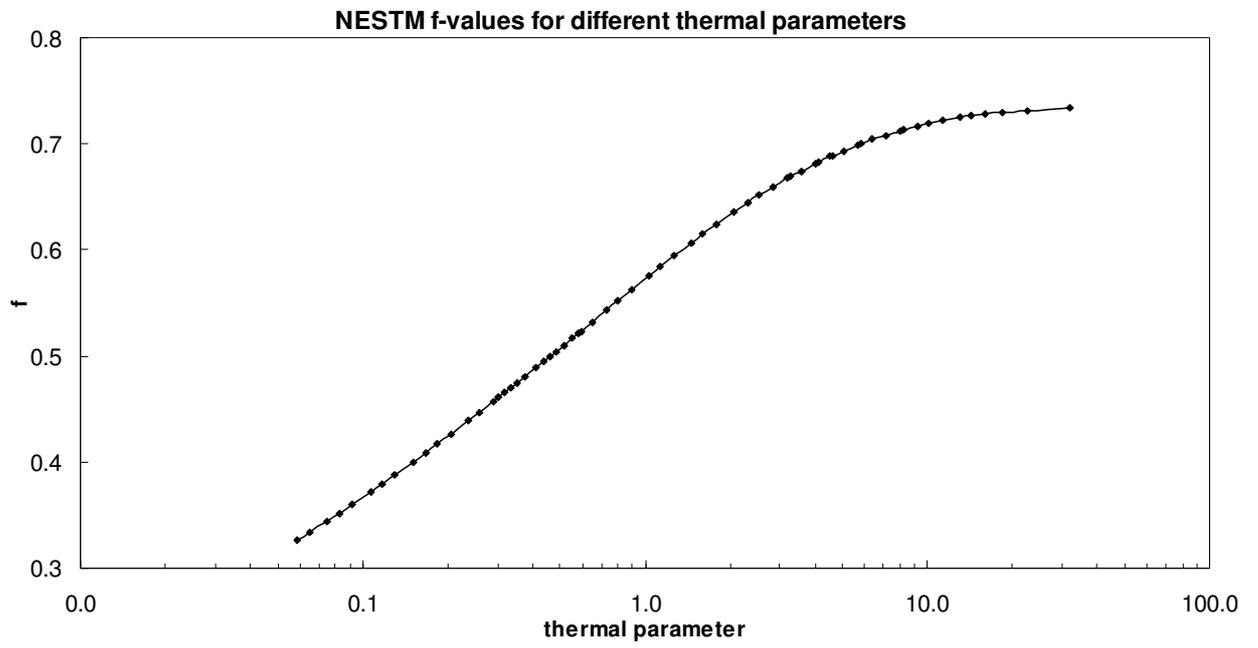

*Fig. 2*

(a)

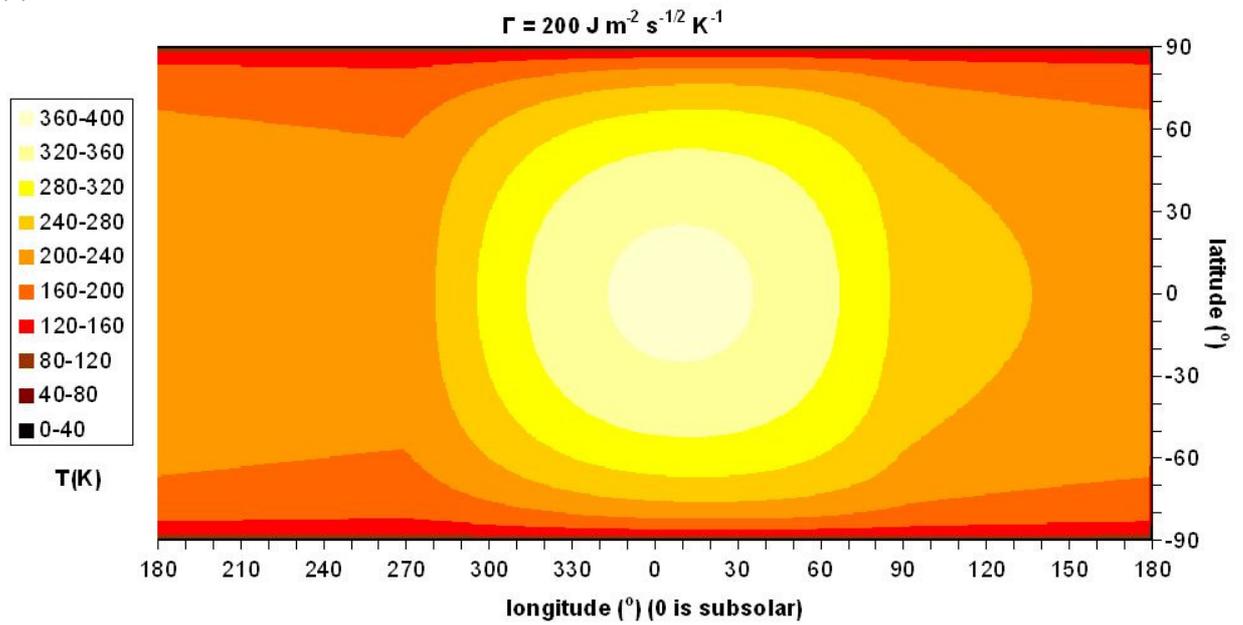

(b)

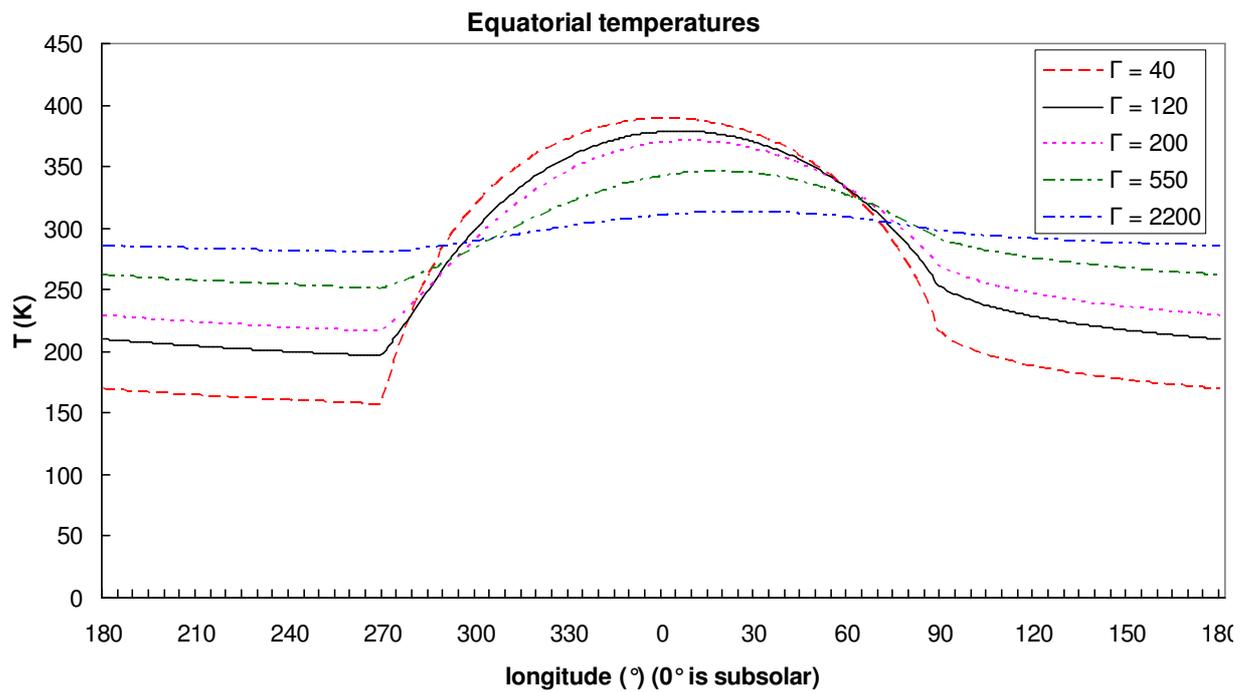

*Fig. 3*
(a)

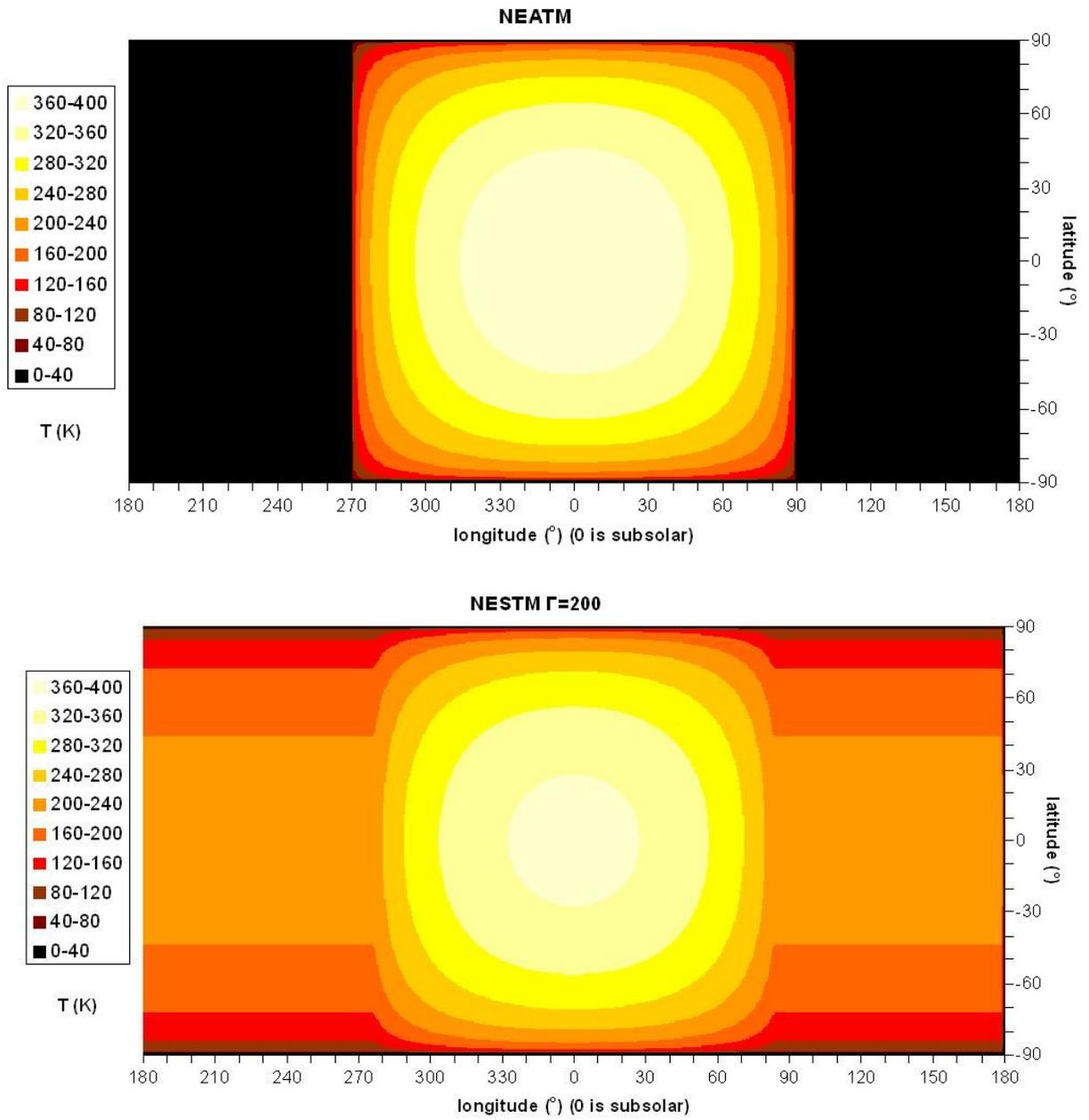

(b)

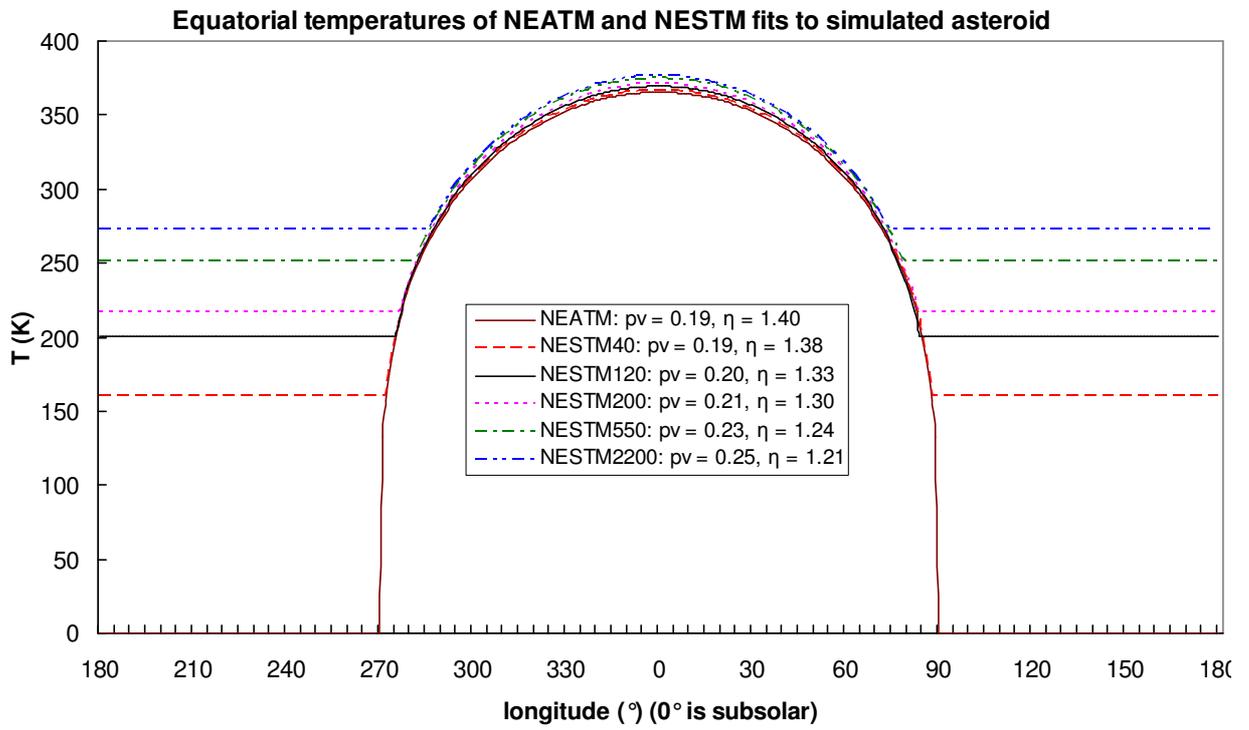

**Equatorial temperatures of NEATM and NESTM fits to simulated asteroid**

Legend:
- NEATM: pv = 0.19, η = 1.40
- NESTM40: pv = 0.19, η = 1.38
- NESTM120: pv = 0.20, η = 1.33
- NESTM200: pv = 0.21, η = 1.30
- NESTM550: pv = 0.23, η = 1.24
- NESTM2200: pv = 0.25, η = 1.21

*Fig. 4*

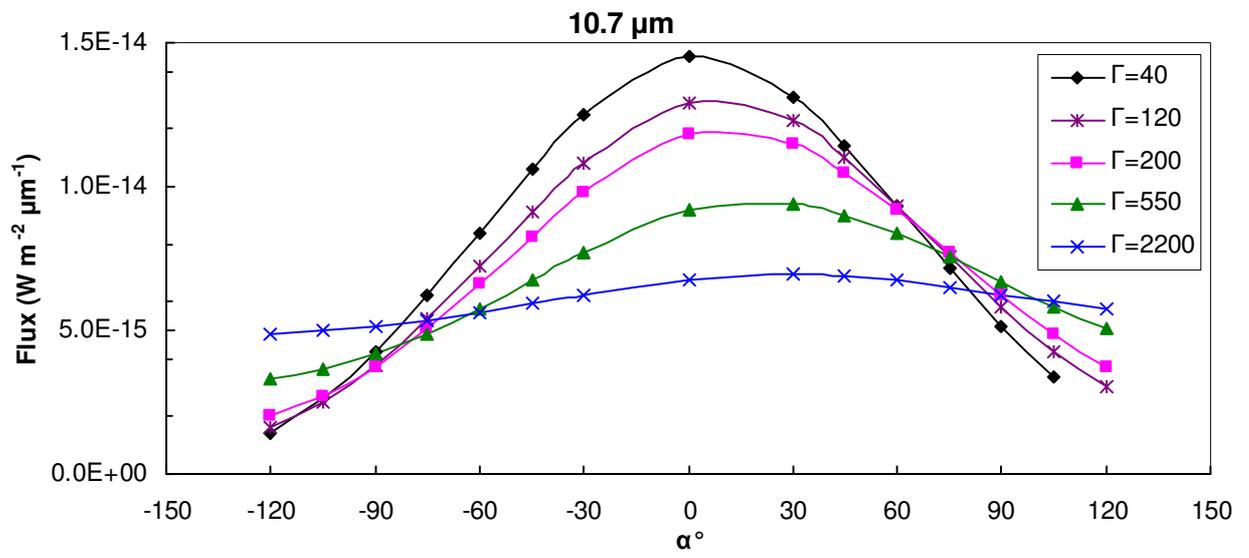

*Fig. 5*
(a)

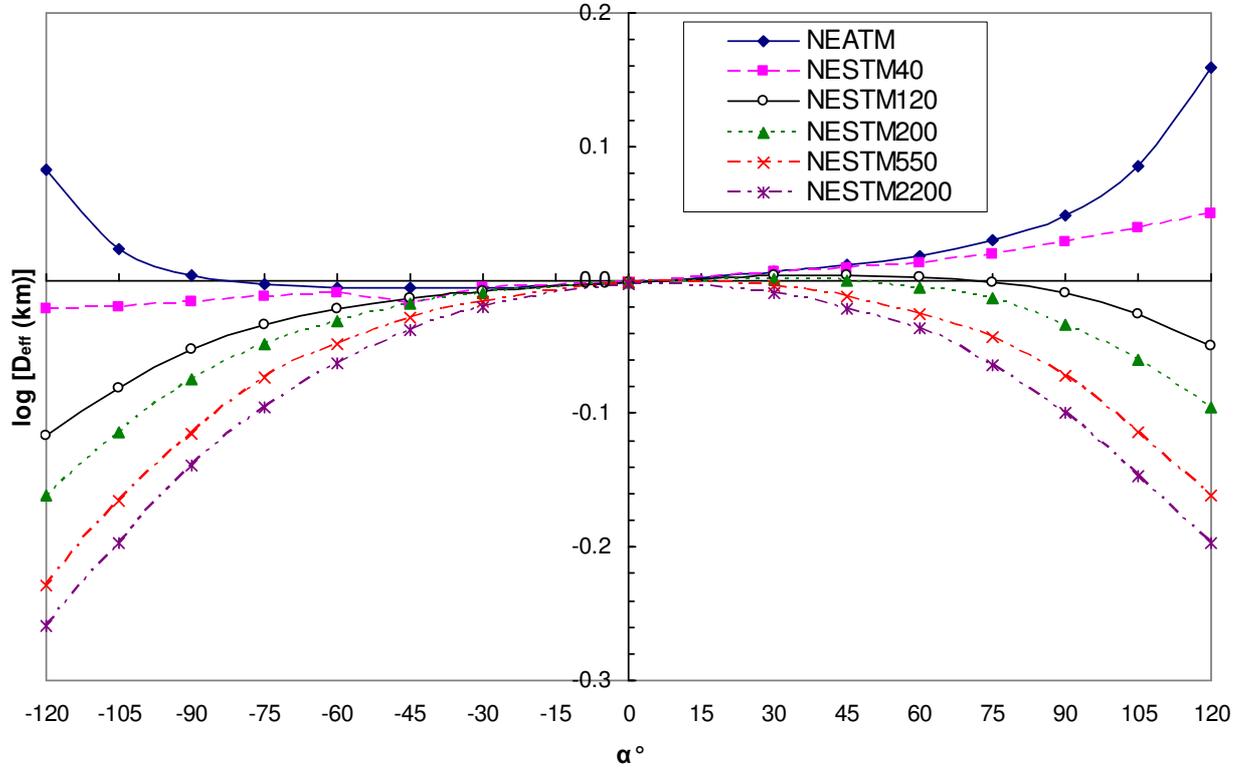

(b)

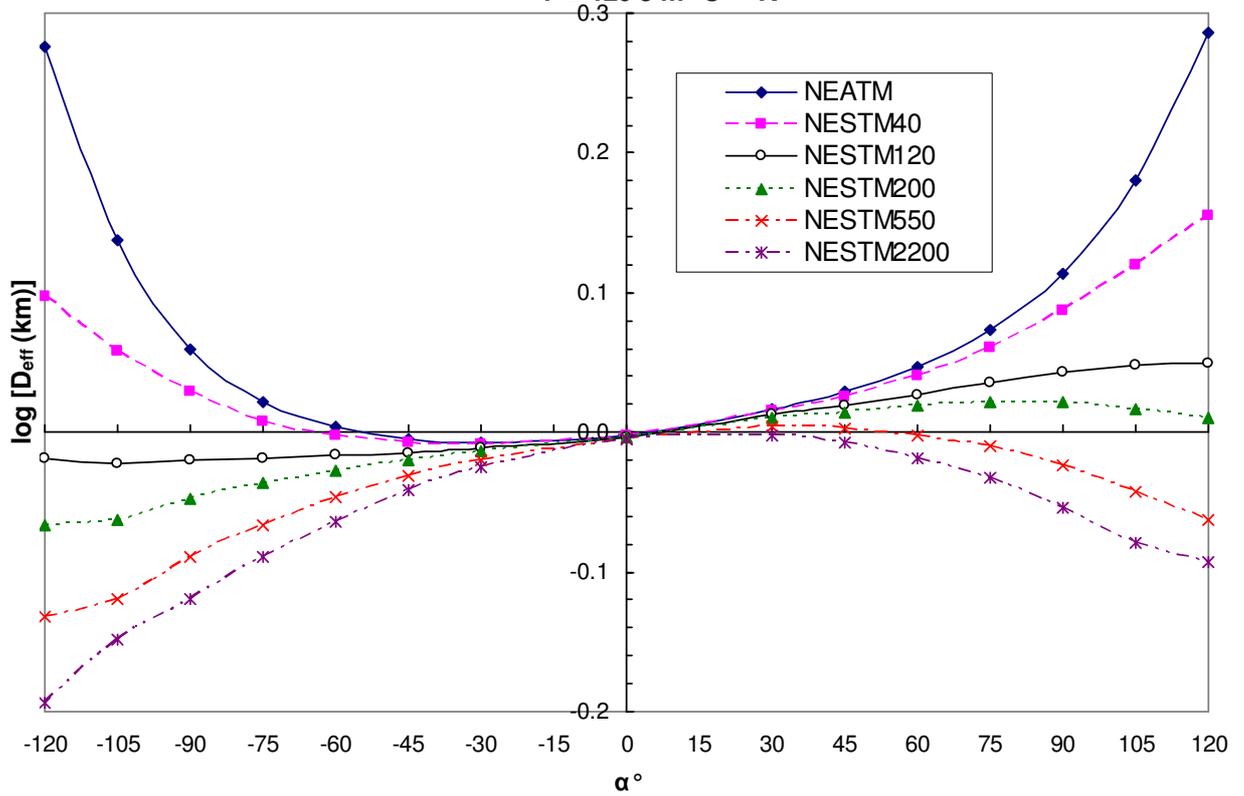

(c)

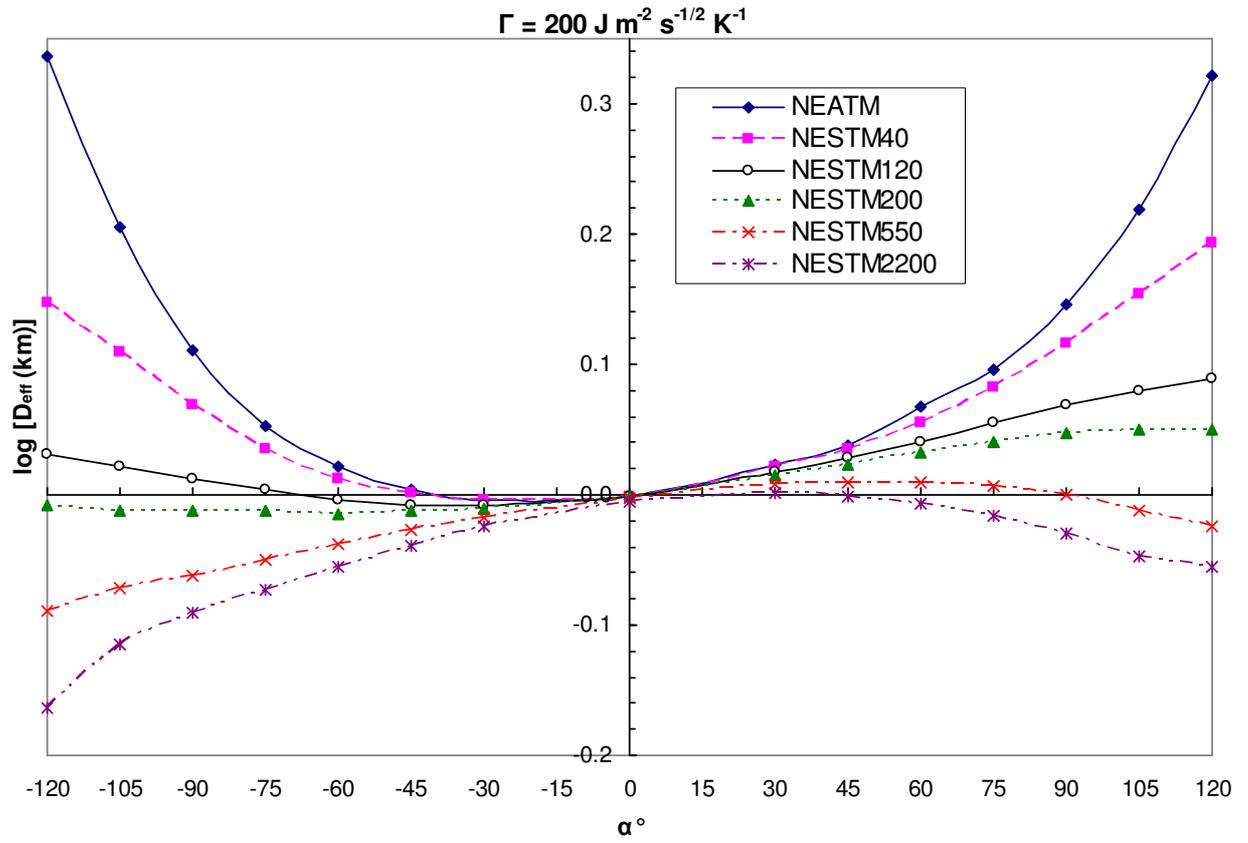

(d)

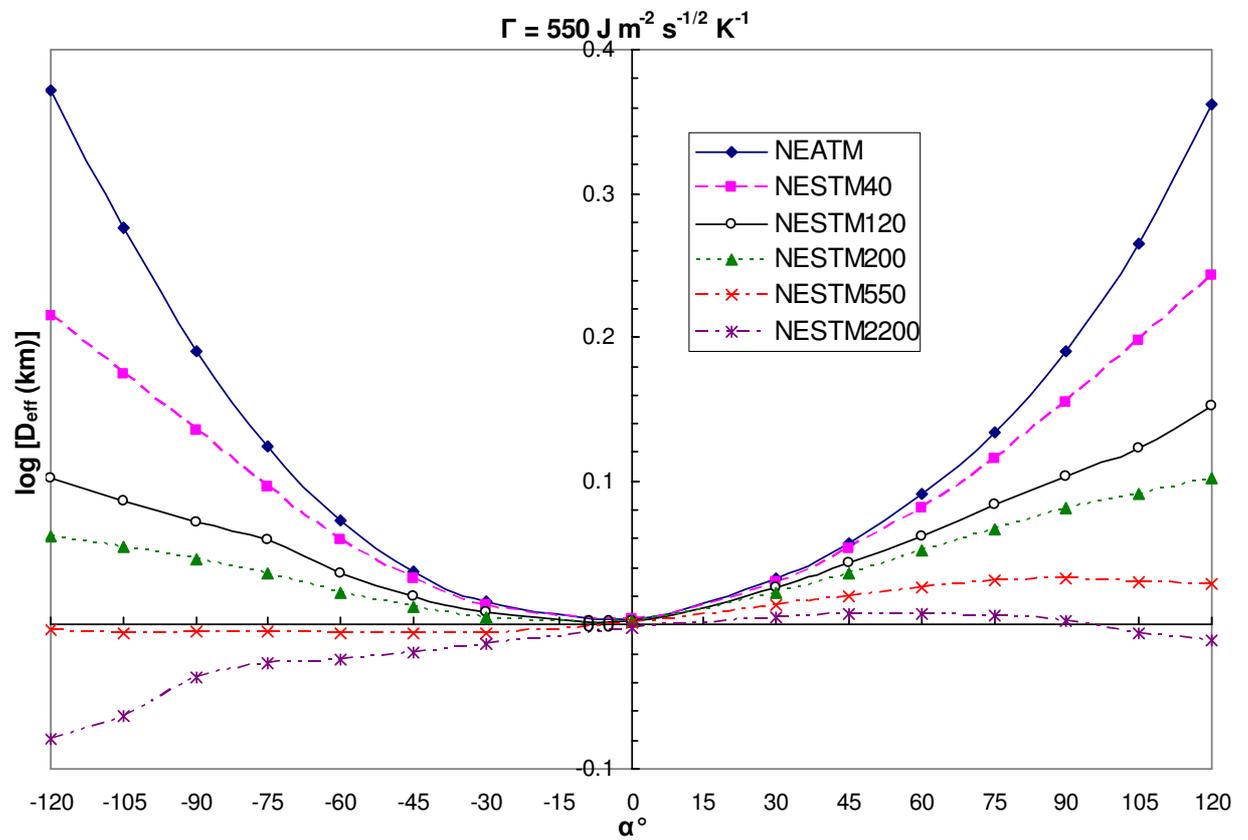

(e)

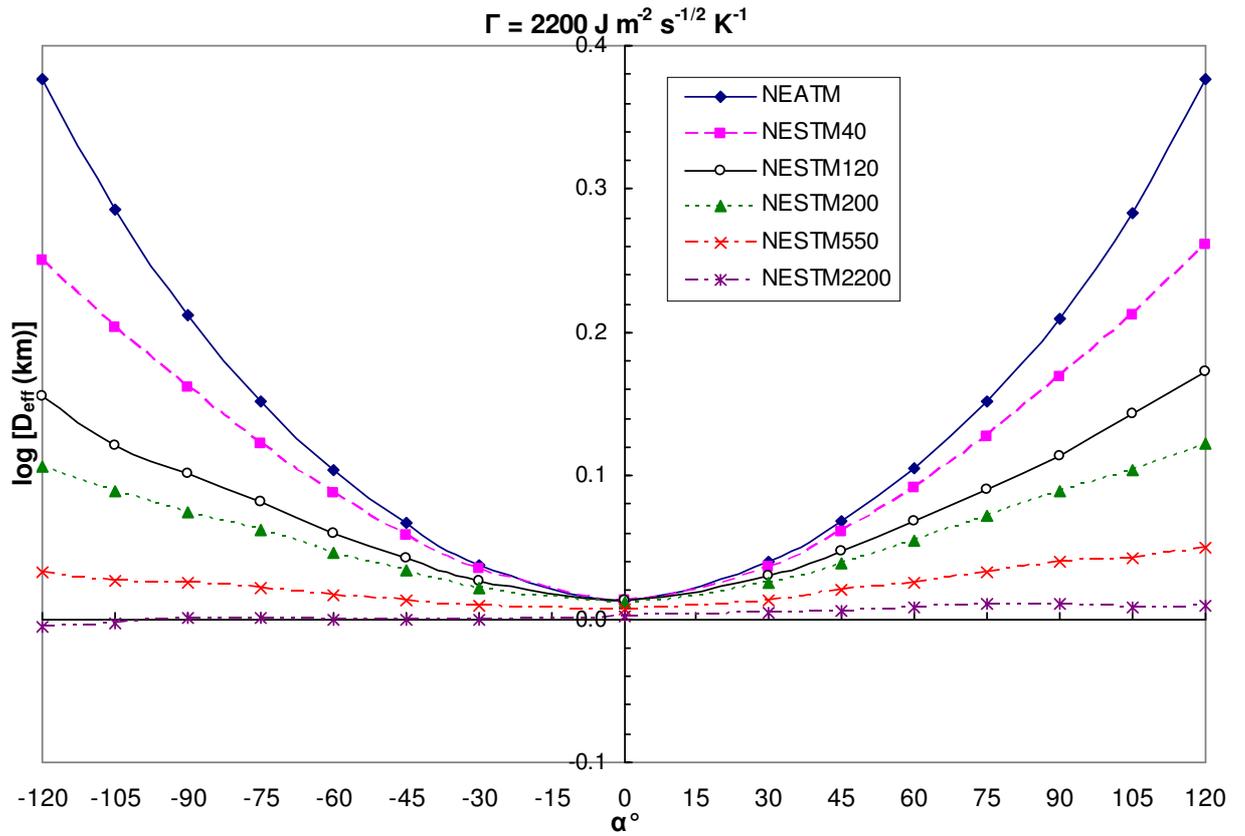

*Fig. 6*

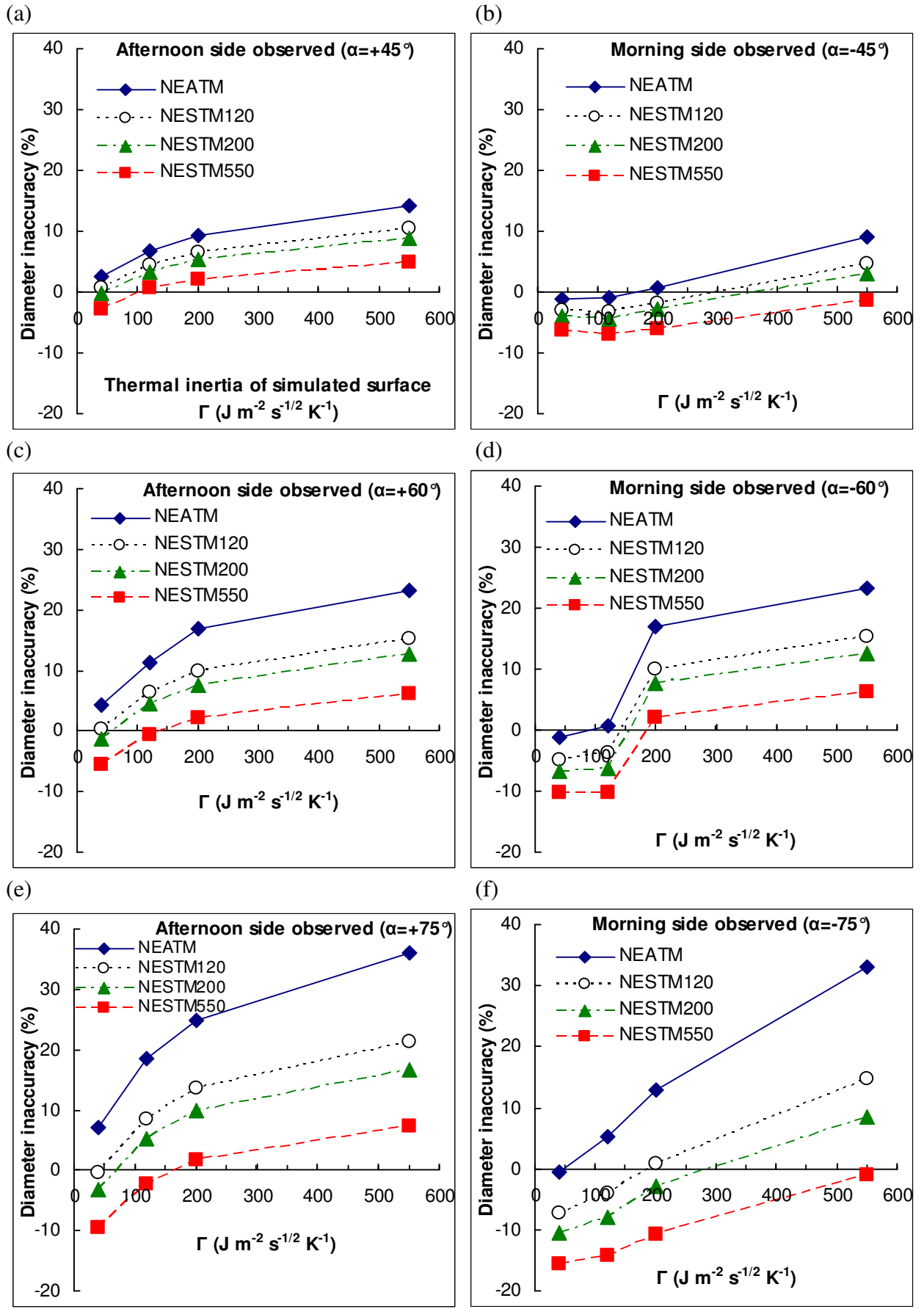

*Fig. 7*

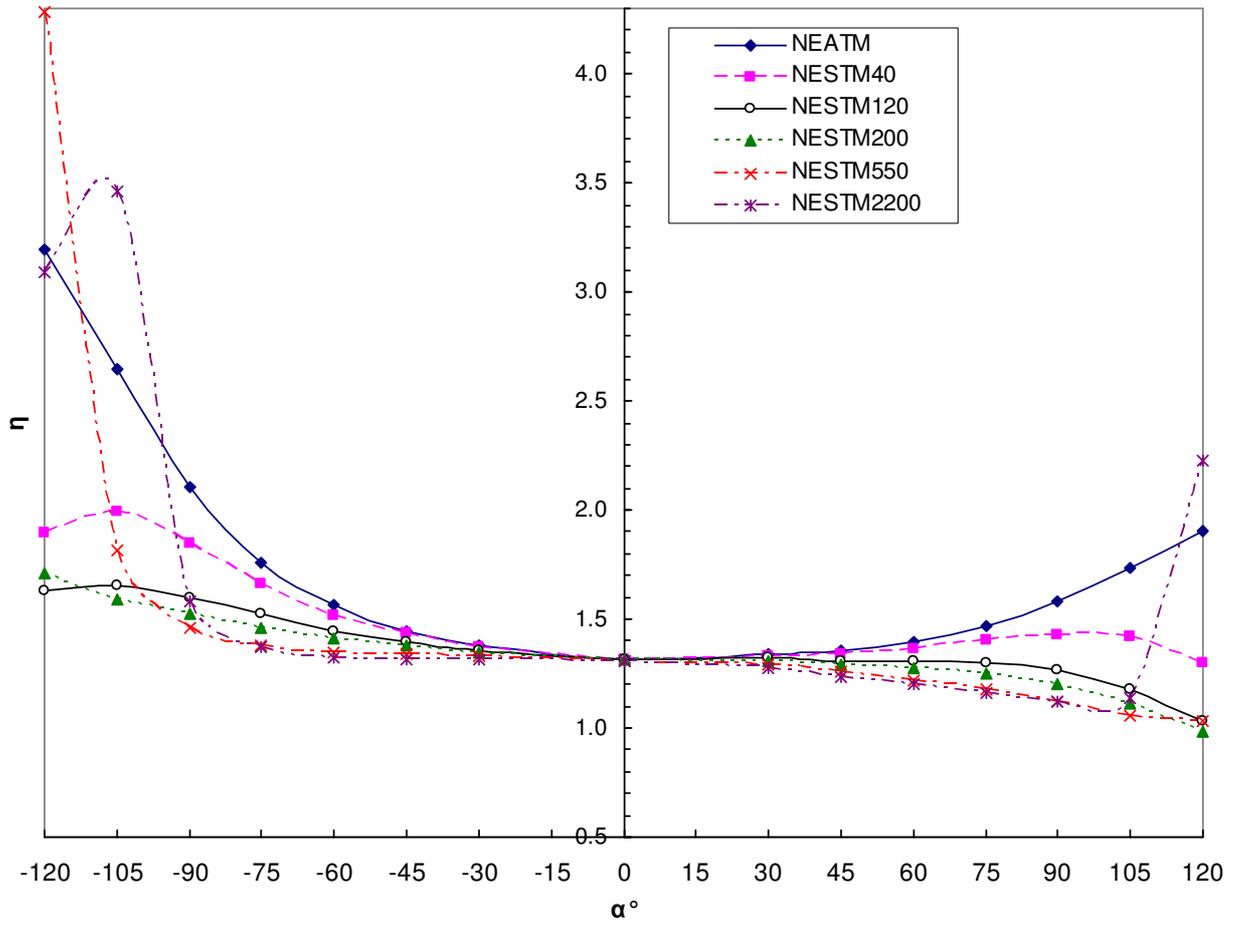

**Variation of best-fit beaming parameter for a smooth 1 km diameter spherical asteroid with surface thermal inertia $\Gamma = 200$ J m$^{-2}$ s$^{-1/2}$ K$^{-1}$**

*Fig. 8*

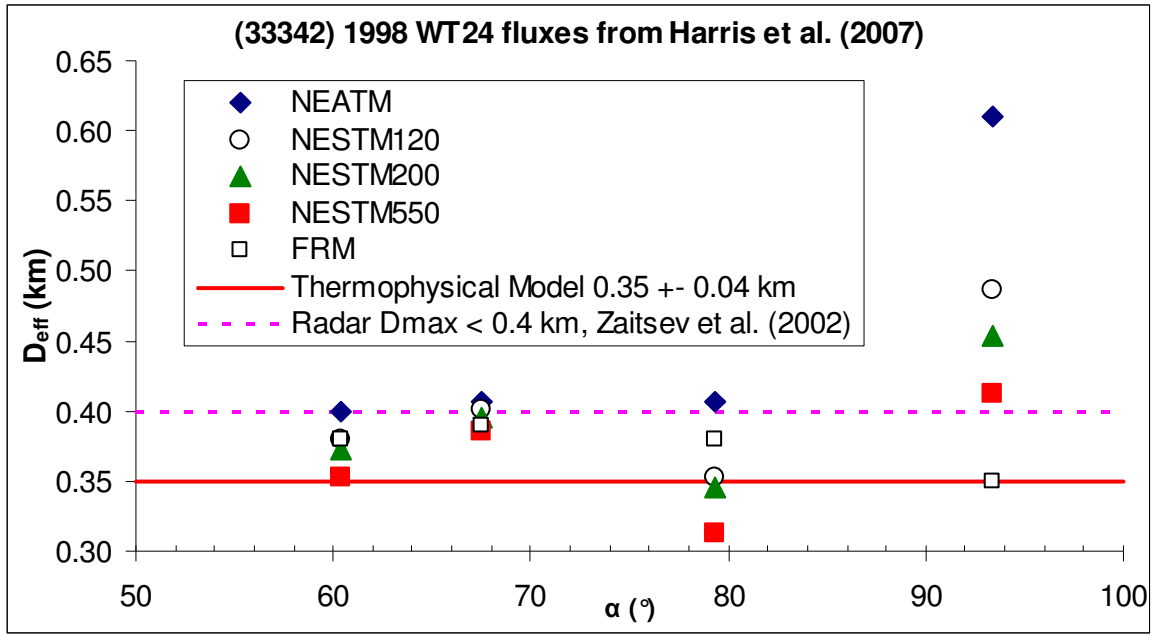

*Fig. 9*

(a)

Relative error between NEATM / NESTM / FRM and radar diameters for observations where η is fitted

(b)

Relative error between NEATM / NESTM / FRM and radar diameters for observations where default η is used

*Fig. 10*

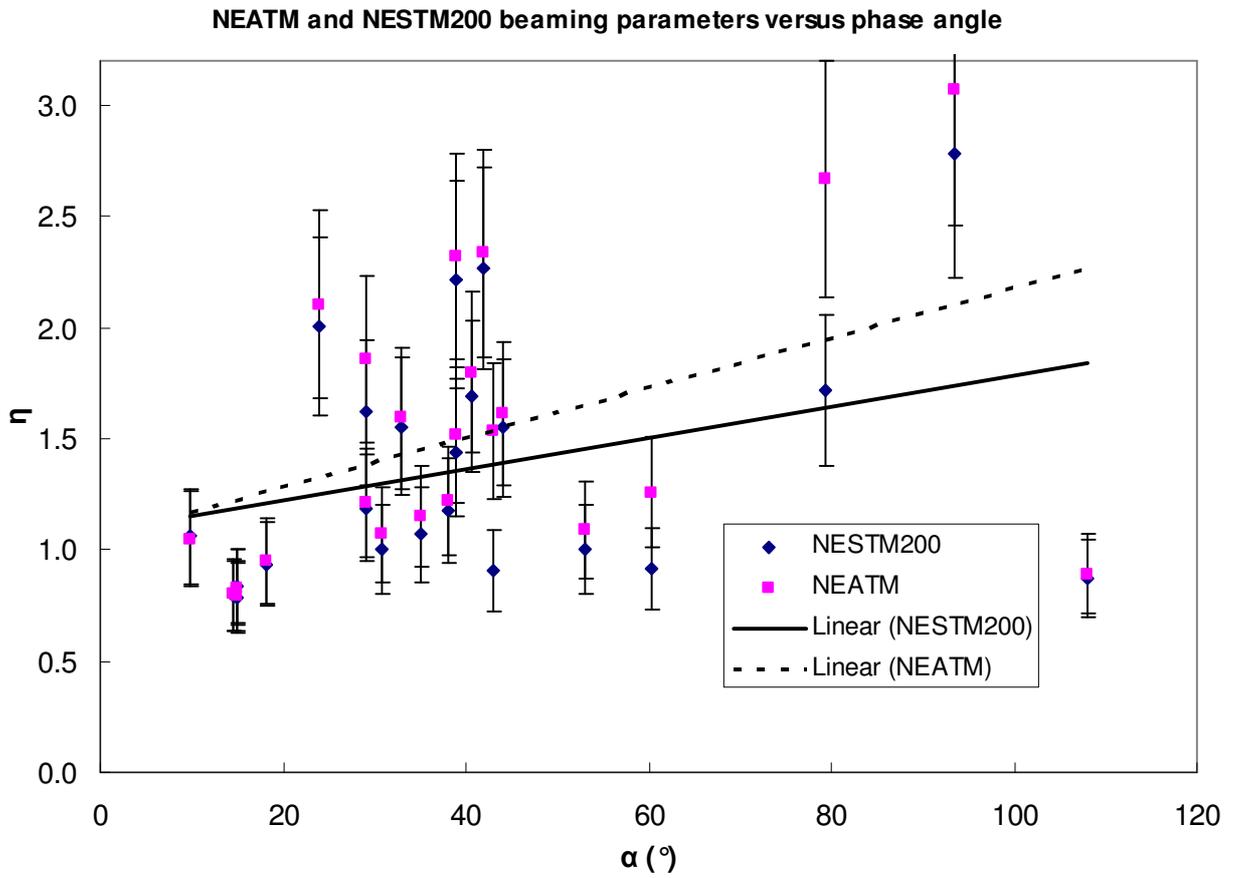

**Figure Captions**

*Fig. 1. f parameters for different thermal parameters $\Theta$ found by ratioing night side equatorial surface temperatures, produced using the thermophysical model, to $T_{max}$.*

*Fig. 2. Surface temperatures derived suing the thermophysical model with A = 0.09815, P = 5h, r = 1 AU: (a) $\Gamma$ =200, (b) equatorial surface temperatures derived for a range of values of thermal inertia. As $\Gamma$ increases, the maximum day side temperature decreases and the night side temperature increases.*

*Fig. 3. Temperature distributions produced by the NEATM and NESTM best-fitting fluxes to a simulated asteroid with $\Gamma$ =200, $p_v$ = 0.25, G = 0.15, r = 1.0 AU, P = 5 h, observed at $\alpha$ = +60°, i.e. on the afternoon side of the asteroid (therefore higher thermal inertia NESTM models fit better). (a) NEATM and NESTM200: for a given latitude there is a constant temperature on the night side. (b) Equatorial temperatures for NEATM, NESTM40, NESTM120, NESTM200, NESTM550 and NESTM2200. These correspond to asteroids with thermal parameter $\Theta$ = 0.234, 0.703, 1.174, 3.251 and 13.081 respectively. The appropriate f parameters are obtained from a look-up table plotted in Fig. 3 and are f = 0.439, 0.543, 0.584, 0.669 and 0.725 respectively. It can be seen that there is zero emission on the night side for NEATM.*

*Fig. 4. Synthetic thermal IR fluxes at 10.7 μm for thermophysical model-derived surface temperatures simulating an asteroid with r = 1.0 AU, P = 5 h and 4 different thermal inertias $\Gamma$, "observed" at a range of different phase angles on the afternoon side (+$\alpha$) and on the morning side (-$\alpha$). Note how for $\Gamma$ =550 and $\Gamma$ =2200 J m$^{-2}$ s$^{-1/2}$ K$^{-1}$ the $\alpha$ = +30° fluxes are actually higher than at $\alpha$ = 0° due to thermal lag (c.f. Fig. 4).*

*Fig. 5. Variation of model diameters with phase angle, observed on the afternoon side (+α) and the morning side (-α), fitting to thermophysical model-derived thermal IR fluxes for an asteroid with $p_v$ =0.25, $D_{eff}$ =1.0 km, P = 5 h at r = 1.0 AU. The NEATM and five different NESTM versions are fitted (resulting in applying f ≈ 0.439, 0.543, 0.584, 0.669 and 0.725 respectively, although f varies depending on the best-fit $p_v$). Asteroid surface with (a) Γ = 40, (b) Γ =120, (c) Γ = 200, (d) Γ = 550, (e) Γ = 2200 J $m^{-2}$ $s^{-1/2}$ $K^{-1}$.*

*Fig. 6. Diameter inaccuracy using NEATM and NESTM for simulated asteroid surfaces with Γ = 40, 120 200, 550 observed at 45, 60° and 75° phase angle on: (a) afternoon side; (b) morning side.*

*Fig. 7. Variation of model best-fit beaming parameters η at different phase angles α, fitting to thermophysical model-derived (with surface Γ = 200 J $m^{-2}$ $s^{-1/2}$ $K^{-1}$) thermal IR fluxes for an asteroid with $p_v$ =0.25, $D_{eff}$ =1.0 km, P = 5 h at r = 1.0 AU. Asteroid is "observed" on the afternoon side (+α) and the morning side (-α). The NEATM and five different NESTM versions are fitted (resulting in using f parameters of approximately 0.439, 0.543 0.584, 0.669 and 0.725 respectively, although f varies depending on the best-fit $p_v$).*

*Fig. 8. NEATM, NESTM and FRM diameters fitted to thermal IR fluxes of (33342) 1998 WT$_{24}$ from Harris et al. (2007) compared with a diameter obtained with a thermophysical model applied to the same data, and to a diameter obtained with radar (Zaitsev et al. 2002).*

*Fig. 9. Comparison of NEATM, NESTM and FRM relative diameter error with radar diameters. (a) Only utilising observations where η could be fitted; (b) only observations where default η*

*was used. Error bars only included for NEATM and FRM for clarity (NESTM uncertainties are similar to NEATM).*

*Fig. 10. Beaming parameter η versus phase angle α for observations fitted with NEATM and NESTM200 given in Tables 2-4. NESTM200 has a shallower trend and lower beaming parameters overall (η = 0.007α + 1.08).*